\documentclass[
preprint,
tightenlines,
superscriptaddress,
 amsmath,amssymb,
 aps,
floatfix,
]{revtex4-1}

\usepackage{xspace}
\usepackage{bm}
\usepackage{graphicx,epsf}
\usepackage{xcolor}
\usepackage{float}
\usepackage{psfrag}
\newcommand{\blue}{}

\begin{document}

\title{Robust topological phase in proximitized core-shell nanowires coupled to multiple superconductors}

\author{Tudor D. Stanescu}
\affiliation{Department of Physics and Astronomy, West Virginia University,
Morgantown, WV 26506, USA}

\author{Anna Sitek}
\affiliation{Department of Theoretical Physics, Faculty of Fundamental
Problems of Technology, Wroclaw University of Science and Technology, Wroclaw, 50-370, Poland}
\affiliation{School of Science and Engineering, Reykjavik University,
Menntavegur 1, IS-101 Reykjavik, Iceland}

\author{Andrei Manolescu}
\affiliation{School of Science and Engineering, Reykjavik University,
Menntavegur 1, IS-101 Reykjavik, Iceland}

\begin{abstract}
We consider core-shell nanowires with prismatic geometry contacted with two or more superconductors in the presence of a magnetic field applied parallel to the wire. In this geometry, the lowest energy states are localized on the outer edges of the shell, which strongly inhibits the orbital effects of the longitudinal magnetic field that are detrimental to Majorana physics. 
Using a tight-binding model of coupled parallel chains, we calculate the topological phase diagram of the hybrid system in the presence of non-vanishing transverse potentials and finite relative phases between the parent superconductors. 
We show that   having  finite relative phases strongly enhances the stability of the induced topological superconductivity over a significant range of chemical potentials and reduces the value of the critical field associated with the topological quantum phase transition.  
\end{abstract}

\maketitle

\section{Introduction}

The intense ongoing search for Majorana zero modes (MZMs) in solid states systems is motivated, in part, by the perspective of using them as a platform for fault-tolerant topological quantum computation \cite{Kitaev01,Kitaev03,Wilczek09,Stanescu2017}.
Several practical realizations of ``synthetic'' topological superconductors that host zero-energy Majorana modes have been proposed in the past few years, the most promising involving semiconductor-superconductor hybrid systems \cite{Sau10,Alicea2010,Oreg10,Lutchyn10,Sau2010}.
The basic idea  \cite{Alicea12,Stanescu13a,Beenakker13,Franz13}  is to proximity-couple a  semiconductor nanowire with strong Rashba-type spin-orbit coupling (e.g., InSb or InAs) to a standard s-type superconductor (e.g., NbTiN or Al)  in the presence of a longitudinal magnetic field. The system is predicted to host zero-energy Majorana modes localized at the two ends of the nanowire \cite{Sau10,Lutchyn10,Oreg10}.  These zero-energy states combine equal proportions of electrons and holes and are created by second quantized operators satisfying the ``Majorana condition'' $\gamma^\dagger = \gamma$. The topological character of these modes endows them with robustness against perturbations that do not close the superconductor gap, e.g.,  weak interactions,  wire bending, a certain amount of disorder, etc. 

The most straightforward experimental signature of a Majorana mode is  a zero-bias conductance peak that is produced in a charge transport measurement by tunneling electrons between the semiconductor wire and external electrodes attached to its ends \cite{Mourik12,Deng12,Das12,Finck13,Churchill13,Albrecht2016,Deng2016,Chen2017,Zhang2017,Nichele2017,Zhang2018}.
These experiments have provided strong indications regarding  the presence of Majorana bound states at the end of the wire, but no clear evidence of a phase transition to the topological phase, as  revealed by the closing of the bulk quasiparticle gap  \cite{Alicea12,Stanescu13a,Beenakker13,Franz13}, or evidence of correlated features at the opposite ends of the wire \cite{DSarma2012}.

Ideally, the MZMs are hosted by a one-dimensional (1D) p-wave
superconductor. However, the experimental realization and detection
of these modes involve 3D nanowires \cite{Lutchyn2017}. The most
common materials are InSb and InAs due to their large g-factor and
strong SOC. The wires are grown by bottom-up methods and have usually
a prismatic shape with a hexagonal cross section, as determined by the
crystal structure \cite{Ihn10}.  The finite cross section of the wires
used in the \blue{experiments} may generate additional phenomena, which are
not captured by ideal 1D models. \blue{In particular, the orbital effects of
the magnetic field, which is oriented  parallel to the nanowire, may
reduce or even destroy the stability of the Majorana modes \cite{Nijholt16}.}

Proximitized core-shell nanowires are slightly more complex systems
recently shown \cite{Manolescu17} to have interesting  Majorana physics
that is practically immune to orbital effects.  With a conductive shell
and an insulating core, such heterostructures become tubular conductors.
\blue{The prismatic shape of the core-shell wires implies that the cross section of
the shell can be seen as a polygonal ring. This is an interesting geometry
because the corners of the polygon act like quantum wells where the states with the lowest energies
are localized. Furthermore, a group of states with higher energies is
localized on the sides of the polygon \cite{Bertoni11}.  Although
most of the core-shell nanowires have a hexagonal profile, square \cite{Hu11}
or triangular \cite{Wong11,Blomers13,Qian12,Heurlin15,Yuan15} cross
sections can also be obtained. The core diameter is typically between 50-500 nm and
the shell thickness is between 1-20 nm.} For all these geometries, the edge states
corresponding to corner localization represent better approximations of
the ideal 1D limit than the states hosted by a full wire.  Remarkably,
the energy separation between the corner states and the side states
increases when the shell thickness is narrow compared to the radius
of the wire, and when the corners are sharp.  This means that the
triangular shell would be the best choice for the realization of 1D
edge channels. For example, with a shell thickens of 8-10 nm and a
radius of 50 nm the energy separation between corner and side states
can be between 50-100 meV \cite{Sitek15,Manolescu17}. In this case the
corner states are extremely robust to orbital effects of the magnetic
field and the low-energy subspace is well separated from higher-energy
states. Another interesting aspect of a prismatic shell is that it can
host several Majorana states at each end of the wire.  One can actually
view the wire as a set of $n$ coupled chains, each having a pair of
Majorana modes at its ends.  On the one hand, this results in a rich
phase diagram \cite{Manolescu17}, which means that core-shell nanowires
provide an interesting playground for studying topological quantum phase
transitions. On the other hand, this richness is associated with rather
fragile topological phases \cite{Manolescu17}. In practice, it would be
extremely useful to have a knob enabling one to control the robustness
of topological superconducting phase.

In this work we show that coupling a core-shell nanowire to two or more parent superconductors with non-vanishing relative phases enhances the stability of the topological phase and lowers the critical magnetic field associated  with the (lowest field) topological quantum phase transition. 	 
\blue{In principle the phase difference between superconductors can be achieved
either by applying an additional magnetic field, i.e., other than the longitudinal field
needed for the Zeeman energy, or by driving a supercurrent 
through the superconductors.}
Hence, by controlling the relative phases of the parent superconductors coupled to the wire one can stabilize the topological superconducting phase that hosts the zero-energy Majorana modes and  one  can obtain  an additional experimental knob for exploring a rich phase diagram and  observing potentially interesting low-energy physics. 

The rest of this article is organized as follows. We first describe the coupled-chains tight binding model that we use in our numerical analysis. Then,  using this simple model, we  study the topological phase diagram of (infinite) core-shell wires with triangular and  square cross section coupled to superconductors having the same superconducting phase. Next, we show that a finite phase difference can stabilize the topological phase in both triangular and square geometries. In addition, we show that the critical field associated with the (low-field) topological quantum phase transition can be made arbitrarily low. The implications of these findings for the stability of the Majorana modes emerging in finite wires is discussed in the subsequent section. Next, we corroborate our results for the topological phase diagram using an alternative ``geometric'' model. Finally, we summarize our findings and present our main conclusions.


\section{\label{toy} The coupled-chains tight-binding model}

\begin{figure}
\begin{center}
\includegraphics[width=0.45\textwidth]{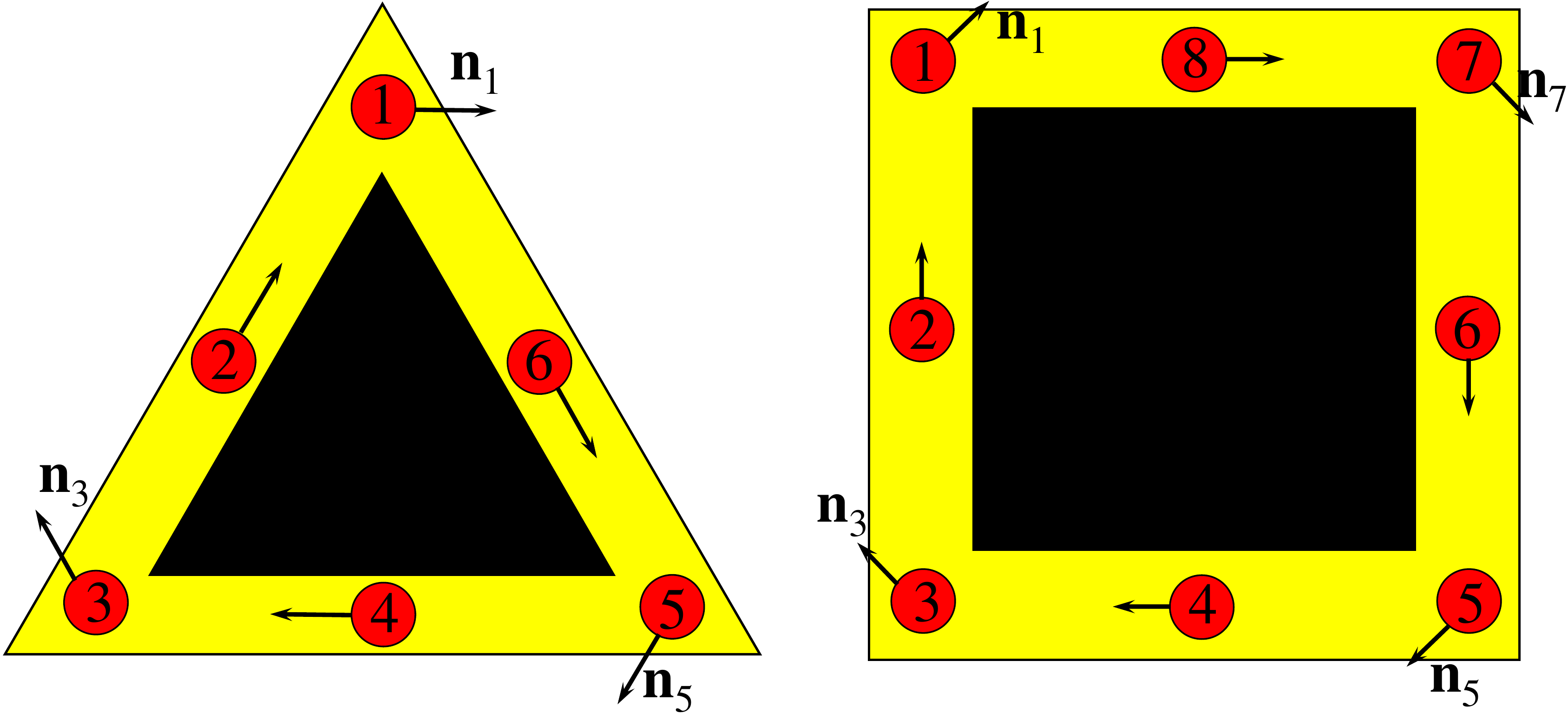}
\end{center}
\caption{Schematic representation of the chain model for triangular (left) and square (right) core-shell nanowires.  The shell (yellow) is coarse-grained so that the vertices and the sides are represented by 1D chains (red circles). The arrows indicate the direction of the effective spin-orbit field ${\bm n}_\ell$ associated with the (longitudinal) Rashba spin-orbit coupling. In a minimal model each side is represented by one chain (left); a more detailed representation can be obtained by adding more chains associated with the sides (right).}
\label{Fig0TT}
\end{figure}

We start by formulating the effective thigh-binding model that describes the low-energy physics of a core-shell nanowire  with $n$ edges.  \blue{The model has already been introduced for triangular core-shell nanowires 
in Ref. \cite{Manolescu17} (Appendix), and also previously considered by other authors, in different forms, for ladder systems \cite{Poyhonen14,Wakatsuki14}.}
A ``coarse-grained'' shell is modeled  by one chain associated with each vertex and one or more chains corresponding to each side, as shown in Fig. \ref{Fig0TT}.
Note that the minimal model for a nanowire with $n$ edges consists of $2n$ coupled chains ($n$ for vertexes and $n$ for sides), but more detailed representations can be obtained by increasing the number of chains associated with the sides.
A model that takes into account the details of the internal geometry of the wire \cite{Manolescu17} will be used later in the paper to corroborate the results obtained with this simple tight-binding model. In the numerical calculations we use minimal tight-binding models consisting of $6$ (for triangular wires) or $8$ (for square wires) parallel chains. Note that the odd chains, $\ell = 1,3,\dots$, correspond to the corners, while the even chains, $\ell = 2,4,\dots$, represent the sides.

Consider now $2n$ 1D coupled chains proximity-coupled to one or more s-wave superconductors. The superconducting proximity effect is incorporated through the pairing potential $\Delta_\ell$, $1\leq \ell \leq 2n$ associated with each chain.  Note that, in principle, the induced pairing
potential may be chain-dependent.  The low-energy physics of the hybrid structure is described by the following Bogoliubov -- de Gennes (BdG) Hamiltonian:
\begin{eqnarray}
H&=&-t\!\sum_{i, \ell, \sigma} \!\left(c_{i+1 \ell \sigma}^\dagger  c_{i \ell \sigma}+h.c.\right)-t^\prime\!\sum_{i, \ell, \sigma} \!\left(c_{i \ell+1 \sigma}^\dagger  c_{i \ell \sigma}+h.c.\right) \nonumber \\
&+&\sum_{i, \ell, \sigma}\left[ V_{\rm eff}(\ell) - \mu\right]c_{i \ell \sigma}^\dagger  c_{i \ell \sigma}+\epsilon_0\sum_{i, \sigma}\sum_{\ell}^{(even)}c_{i \ell \sigma}^\dagger  c_{i \ell \sigma} \label{eq1T} \\
&+&\frac{i}{2}\sum_{i, \ell}\left[\alpha~\! c_{i+1 \ell}^\dagger \left(\hat{\bm{\sigma}}\cdot {\bm{n}}_\ell\right) c_{i \ell} +\alpha^\prime ~\! c_{i \ell+1}^\dagger  \hat{\sigma}_z c_{i \ell} +h.c.\right] \nonumber \\
&+& \Gamma_B \sum_{i, \ell} c_{i \ell}^\dagger  \hat{\sigma}_z c_{i \ell} + \sum_{i, \ell} \left(\Delta_\ell c_{i \ell \uparrow}^\dagger c_{i \ell \downarrow}^\dagger + \Delta_\ell^*c_{i \ell \downarrow}  c_{i \ell \uparrow} \right), \nonumber
\end{eqnarray}
where $c_{i \ell \sigma}$ is the annihilation operator for an electron with spin projection $\sigma$ localized on the lattice site $i$ of
the chain $\ell$ and $c_{i \ell}=\left(c_{i \ell \uparrow}, c_{i \ell \downarrow}\right)^T$ is the corresponding spinor operator. The first two terms
in Eq. (\ref{eq1T}) represent the nearest-neighbor hopping along the chains, with characteristic energy $t$, and  the inter-chain coupling, with characteristic energy $t'$. In the summations over the chain index $\ell$ we use the convention $2n+1\equiv 1$. 
The third term of the Hamiltonian (\ref{eq1T}) contains a chain-dependent effective potential $V_{\rm eff}(\ell)$ that incorporates the presence of various external electrostatic fields (e.g., gate potentials) and the chemical potential $\mu$. Note that, in general,  $V_{\rm eff}(\ell)$  breaks the $n$-fold rotation symmetry of  the original nanowire.  The term proportional to $\epsilon_0$ accounts for the fact that the side states have higher energies than the corner states and the parameter   $\epsilon_0>0$ controls the energy gap between the two types of states. The next term represents the  Rashba type spin-orbit coupling (SOC), with longitudinal and transverse components proportional to $\alpha$ and $\alpha^\prime$, respectively. The underlying assumption is that the spin-orbit coupling is generated by an effective potential in the shell region due to the presence of the core \cite{Manolescu17}. The corresponding  direction of the spin-orbit field ${\bm n}_\ell$ for electrons moving along the wire is shown in Fig. \ref{Fig0TT}.    The next term in Eq. (\ref{eq1T}), $\Gamma_B= g \mu_b B$,  corresponds to the Zeeman spin splitting generated by an external  magnetic field applied parallel to the wire (e.g., along the z-axis).  The last term describes the proximity-induced pairing and takes into account the possibility that pairing potential $\Delta_\ell$ be chain-dependent.  We assume that the vertex regions are covered by $n$ different superconductors separated by gaps over the side regions. The corresponding proximity-induced pairing potentials are
\begin{equation}
\Delta_{\ell} =\left\{
\begin{array}{l}
0~~~~~~~~~\mbox{if}~\ell~\mbox{is even}, \\
\Delta e^{i\phi_\ell}~~~\mbox{if}~\ell~\mbox{is odd},
 \end{array}\right.                                                                                \label{eq2T}
\end{equation} 
where $\phi_\ell$, the phase of the superconductor coupled to the vertex $\ell$, is an experimentally-controllable quantity. In the numerical
calculations presented below we use the following values for the model parameters: $t=5.64~$meV, $t^\prime=1.41~$meV (or $t^\prime=2.25~$meV, when explicitly specified), $\alpha=2.0~$meV, $\alpha^\prime=0.5~$meV, $\epsilon_0=15.0~$meV, and $\Delta=0.3~$meV.

\blue{
To determine whether a given superconducting phase is topologically trivial or not, we calculate the ${\mathbb Z}_2$ topological
index ${\mathcal M}$, i.e., the {\em Majorana number} \cite{Kitaev01},
\begin{equation}
{\mathcal M} = {\rm sign}\left[{\rm Pf}~\! B(0)\right]{\rm sign}\left[{\rm Pf}~\! B(\pi)\right].
\label{Pff}
\end{equation}
The trivial and topological superconducting phases are characterized by  ${\mathcal M} = +1$ and ${\mathcal M} = -1$, respectively.
In Eq.\ (\ref{Pff}) ${\rm Pf}[\dots]$ represents the Pfaffian \cite{Wimmer2012}, while the antisymmetric matrix $B(k)$  is the Fourier transform of the Hamiltonian (\ref{eq1T}) in the Majorana basis. The matrix $B(k)$ can be constructed using the particle-hole symmetry of the BdG Hamiltonian \cite{Lutchyn10,Ghosh2010},
\begin{equation}
{\mathcal T}{\mathcal H}(k){\mathcal T}^{-1} = {\mathcal H}(-k),   \label{TR}
\end{equation}
where ${\cal H}(k)$ is the Fourier transform of the (single particle) Hamiltonian corresponding to Eq. (\ref{eq1T}) and ${\mathcal T} = U_t K$ is the
antiunitary time-reversal operator, with $U_t$  a unitary operator and $K$ the complex conjugation. Explicitly, we have
\begin{equation}
B(\Lambda) = {\mathcal H}(\Lambda)U_t,     \label{Bmatrix}
\end{equation}
where $\Lambda=0, \pi/a$ are the time-reversal invariant points characterized by the property ${\mathcal H}(-\Lambda)={\mathcal H}(\Lambda)$. The antisymmetry of $B(k)$ at the time-reversal invariant points, $B^T(\Lambda) = -B(\Lambda)$, is a direct consequence of  Eqs. (\ref{TR}) and (\ref{Bmatrix}). Taking into account that for typical parameter values the Pfaffian is always positive at the boundary of the Brillouin zone,
${\rm sign}\left[{\rm Pf}~\! B(\pi)\right]=+1$,  we conclude that the topological phase boundary is determined by a sign change of
${\rm Pf}~\! B(0)$. Finally, using the general relation between the Pfaffian of a skew matrix $A$ and its determinant, $[{\rm Pf}(A)]^2 ={\rm Det}(A)$,
we have ${\rm Det} {\mathcal H}(0) = \left[{\rm Pf}~\! B(0)\right]^2$. Note that ${\rm Det} {\mathcal H}(0)=0$ signals the presence of gapless states.  Thus, the phase boundary, which corresponds to a sign change of the Pfaffian, is accompanied by the closing of the quasiparticle gap at $k=0$ .
}

\section{Nanowire coupled to superconductors with no relative phase difference}

\begin{figure}
\begin{center}
\includegraphics[width=8.0cm]{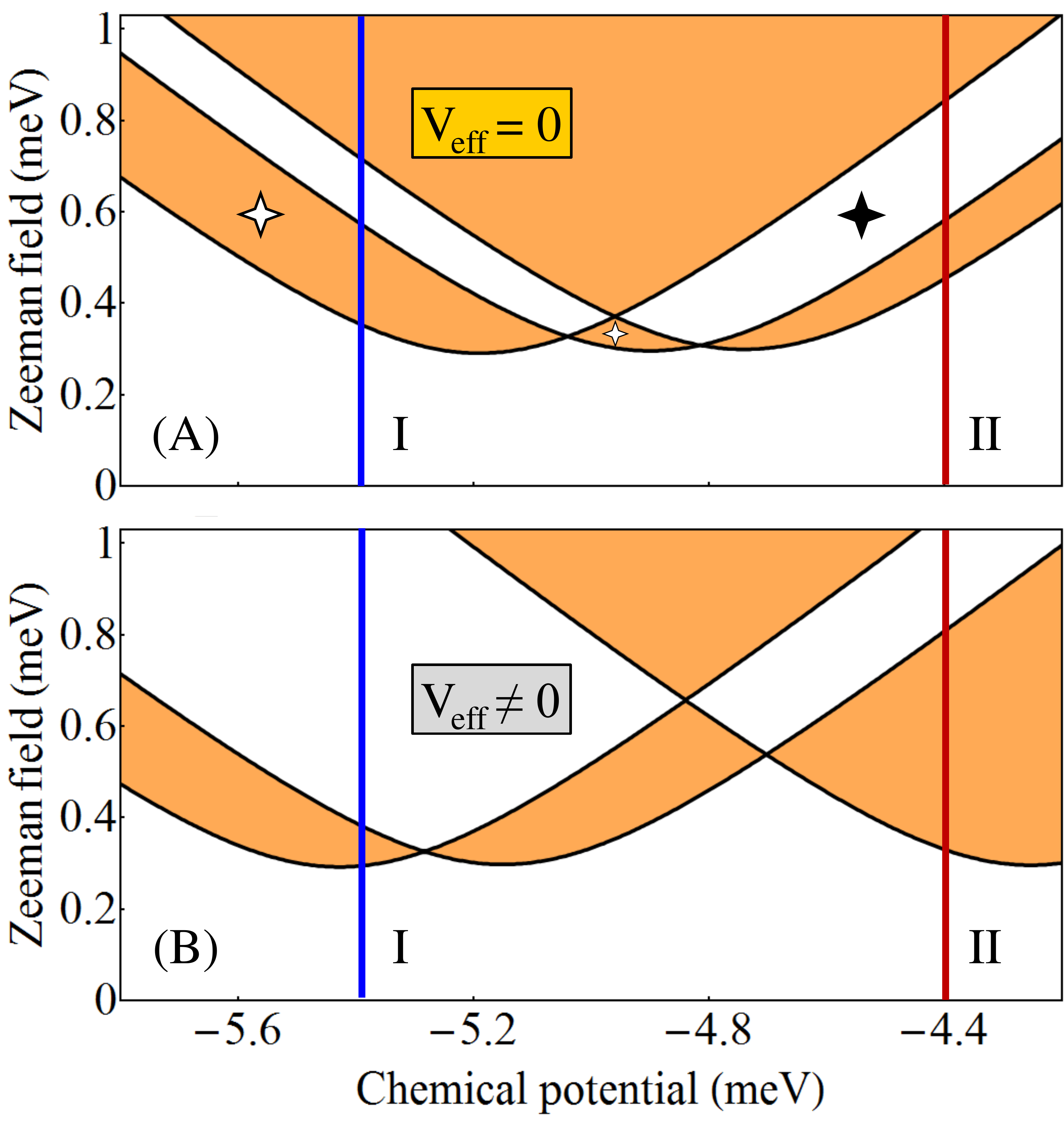}
\end{center}
\caption{(A) Topological phase diagram for a triangular wire with $V_{\rm eff}(\ell)=0$ and $\phi_\ell=0$. The white areas are topologically trivial and the orange regions are nontrivial. The 4-star symbols indicate gapless superconducting phases. (B) Topological phase diagram for a triangular wire with $V_{\rm eff}(\ell)\neq 0$ and $\phi_\ell=0$. The values of the effective potential on the 6 chains are $ (0.67, 0.17, -0.33, -0.33, -0.33, 0.17)$ meV.
The evolution of the (minimum) quasiparticle gap along the cuts I (blue lines) corresponding to $\mu=-5.4~$meV and II (red lines)  corresponding to $\mu=-4.4~$meV are shown in Fig. \ref{Fig2TT} and Fig. \ref{Fig3TT}, respectively.
\blue{See also Ref. \cite{Manolescu17}}
} 
\label{Fig1TT}
\end{figure}

The emergence of topological superconductivity and zero-energy Majorana bound states in core-shell nanowires coupled to a single superconductor (i.e. in the absence of superconducting phase differences)  was discussed in Ref. \cite{Manolescu17}. Here, we summarize the main results, as revealed by the  simplified tight-binding model given by Eq. (\ref{eq1T}).  First, we consider a triangular system without a symmetry-breaking potential,
$V_{\rm eff}(\ell)=0$, and no superconducting phase difference, $\phi_\ell=0$. The corresponding topological phase diagram (as function of the chemical potential and the applied Zeeman field) is  shown in panel (A) of Fig. \ref{Fig1TT}.  The white regions correspond to ${\mathcal M}=+1$ ( i.e. topologically trivial phases), while the orange areas represent topologically nontrivial phases with ${\mathcal M}=-1$. 
The effect of a symmetry-breaking potential is illustrated in  panel (B) of Fig. \ref{Fig1TT}. While the topology of the phase diagram is the same, the phase boundaries are modified significantly with respect to panel (A). We note that this result was  obtained by applying a rather modest symmetry breaking potential with values $V_{\rm eff} = (0.67, 0.17, -0.33, -0.33, -0.33, 0.17)$ meV on the six chains. 

\begin{figure}
\begin{center}
\includegraphics[width=8.0cm]{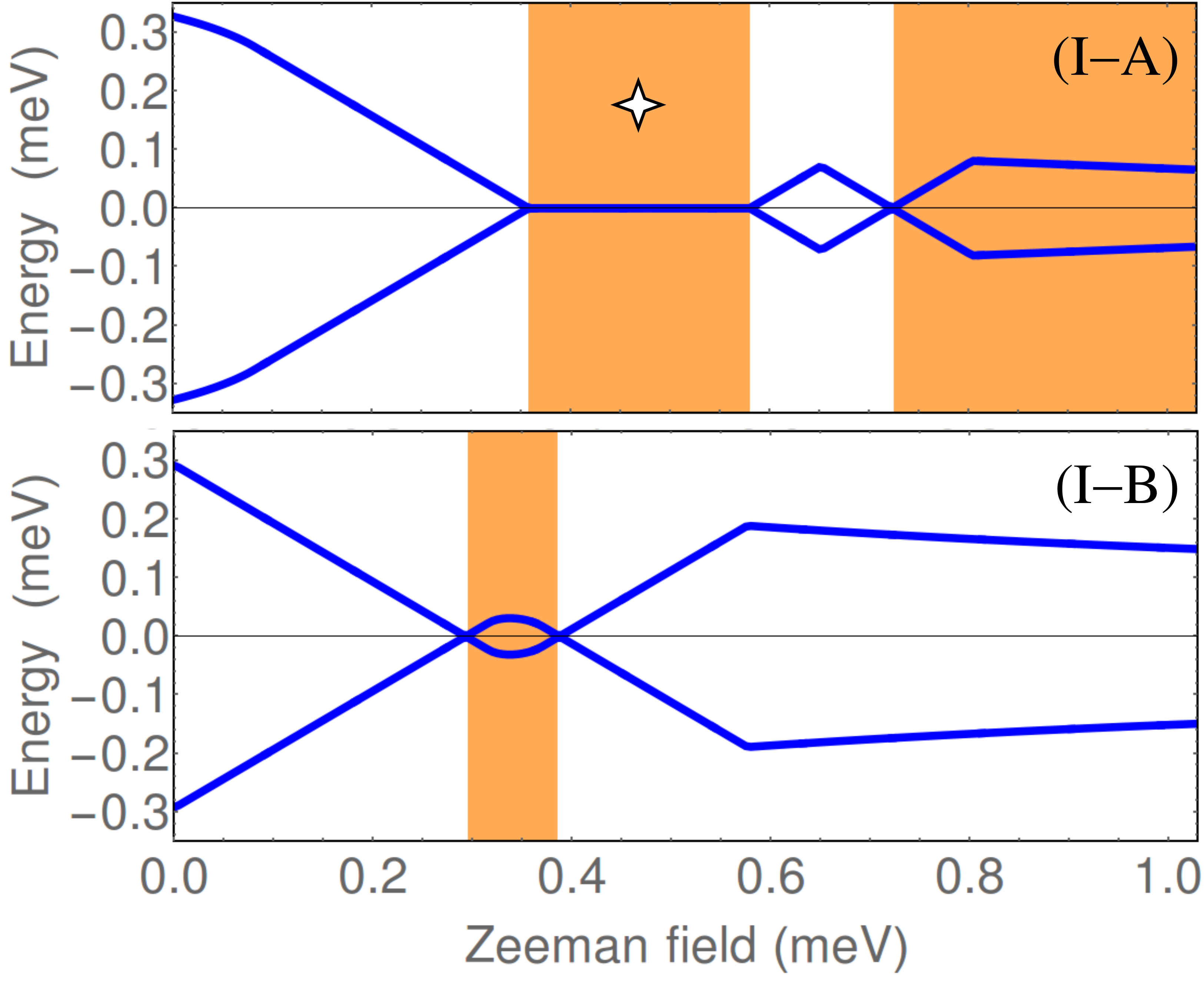}
\end{center}
\caption{Dependence of the minimum quasiparticle gap on the Zeeman field along the blue cuts (I) corresponding to $\mu=-5.4~$meV in Fig. \ref{Fig1TT}. {\em Top}: $V_{\rm eff}(\ell)=0$, see Fig. \ref{Fig1TT}(A). {\em Bottom}:  $V_{\rm eff}(\ell)\neq 0$, see Fig. \ref{Fig1TT}(B). The
white/orange regions correspond to the trivial/nontrivial phases shown in  Fig. \ref{Fig1TT}. Note the gapless superconducting phase marked be the 4-star symbol (top panel).
\blue{See also Ref. \cite{Manolescu17}}}
\label{Fig2TT}
\end{figure}

\begin{figure}
\begin{center}
\includegraphics[width=8.0cm]{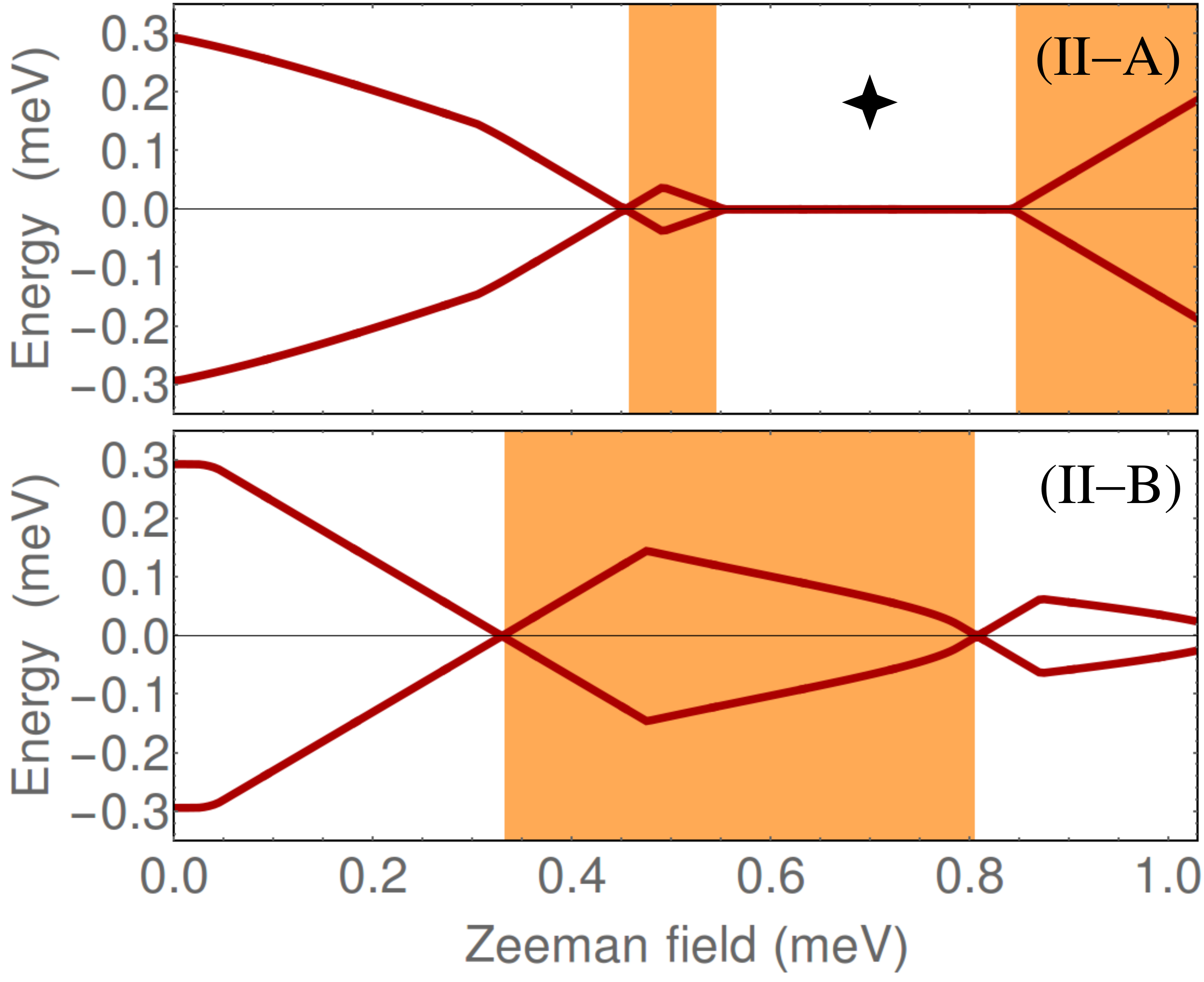}
\end{center}
\caption{Dependence of the minimum quasiparticle gap on the Zeeman field along the dark red cuts (II) corresponding to $\mu=-4.4~$meV in Fig. \ref{Fig1TT}. {\em Top}: $V_{\rm eff}(\ell)=0$, see Fig. \ref{Fig1TT}(A). {\em Bottom}:  $V_{\rm eff}(\ell)\neq 0$, see Fig. \ref{Fig1TT}(B). The
white/orange regions correspond to the trivial/nontrivial phases shown in  Fig. \ref{Fig1TT}. Note the gapless superconducting phase marked be the 4-star symbol (top panel).
\blue{See also Ref. \cite{Manolescu17}}}
\label{Fig3TT}
\end{figure}

To get further insight into the nature of the phases shown in Fig. \ref{Fig1TT}, we  calculate the minimum quasiparticle energy $E_{min}(\mu,\Gamma_B)$ along the constant chemical potential cuts I (blue) and II (dark red) marked on the phase diagrams. This energy (which corresponds to the minimum quasiparticle gap) is defined as
\begin{equation}
E_{min}(\mu,\Gamma_B) = \pm {\rm min}_{n, k}|E_n(k)|,
\end{equation}
where $E_n(k)$ are the eigenvalues of the BdG Hamiltonian from Eq. (\ref{eq1T}). The dependence of $E_{\rm min}$ on the Zeeman field for $\mu=-5.4$ (i.e. the blue cuts I in Fig.\ \ref{Fig1TT}) is shown in Fig. \ref{Fig2TT}, while the evolution of the minimum gap along the cuts II (dark red) corresponding to $\mu=-4.4$ is shown in Fig. \ref{Fig3TT}.


At zero Zeeman field, $\Gamma_B=0$, the system is in a trivial
superconducting phase characterized by a quasiparticle gap
$\Delta=0.3~$meV (see Figs. \ref{Fig2TT} and  \ref{Fig3TT}) given by
the value of the induced pairing potential. With increasing $\Gamma_B$,
the quasiparticle gap reduces and eventually closes at a certain critical
Zeeman energy. In the absence of a symmetry breaking potential, the system
with $\mu=-5.4~$meV [see cut (I-A) in Fig. \ref{Fig1TT}] remains gapless
throughout the first (i.e. low-field) orange region, which means that the
system  becomes a gapless superconductor.  Another gapless superconducting
phase corresponds to the intermediate white region in panel (II-A) of
Fig. \ref{Fig3TT}, i.e. for Zeeman fields between approximately $0.55~$meV
and $0.85~$meV.  These gapless phases are marked by a 4-star symbol in the
phase diagram [see Fig. \ref{Fig1TT}(A)] and in Figs. \ref{Fig2TT}(I-A)
and  \ref{Fig3TT}(II-A). \blue{We note that inside the gapless superconducting
phases the gap closes at $k\neq 0$.  Of course, at the phase boundaries the gap 
always closes at $k=0$.  Furthermore, by increasing the Zeeman energy above 
0.7~meV in panel (I-A) of Fig. \ref{Fig2TT} or above $0.85~$meV
in panel (II-A) of Fig. \ref{Fig3TT}, the system evolves into topological phase with 
a finite gap.}

Upon breaking the three-fold rotation symmetry of the original triangular wire, the gapless superconducting phases become gapped.
Also notice in panel (II-B) that the low-field topological phase corresponding to $\mu=-4.4~$meV is now characterized by a sizable
quasiparticle gap, indicating a regime which may be more favorable for robust zero-energy  Majorana modes. We note that the robust low-field topological phase in panel (II-B) corresponds to a single pair of Majorana modes (i.e. one MZM at each end of the wire) hosted by chain 1 (with the highest value of $V_{\rm eff}$, while the narrow low-field topological phase in panel (I-B) corresponds to a  pair of Majorana modes shared by chains 2 and 3 (the chains with the lowest value of the potential). 
\blue{
Note that the expression ``hosted by chain 1'' (or chains 2 and 3)  actually
means that most of spectral weight associated with the Majorana wave 
function is localized on the corresponding chain(s) (also see below,
Figs. \ref{Fig2exTT} and \ref{Fig4exTT}].
}
The wide trivial region above $\Gamma_B\approx 0.4~$meV in panel (I-B) corresponds to a finite system with two pairs of Majorana bound states (on chains 2 and 3). We also note that the low-field  phase boundaries converge to a single boundary in the
limit of isolated chains, i.e. when the inter-chain hopping energy is much smaller than the hopping along the chains, $t'/t \to 0$.  In this case
three Majorana pairs would form independently at the ends of each chain, and coexist at zero energy, without ``talking" to each other.  Physically, the limit $t'/t \to 0$ corresponds an infinitely-thin shell. For finite  values of $t'/t$  (corresponding to finite shell thicknesses), the coupling between chains lifts the degeneracy, such that at most one Majorana state can have zero energy, while the other two will acquire finite energy.

\begin{figure}
\begin{center}
\includegraphics[width=16.0cm]{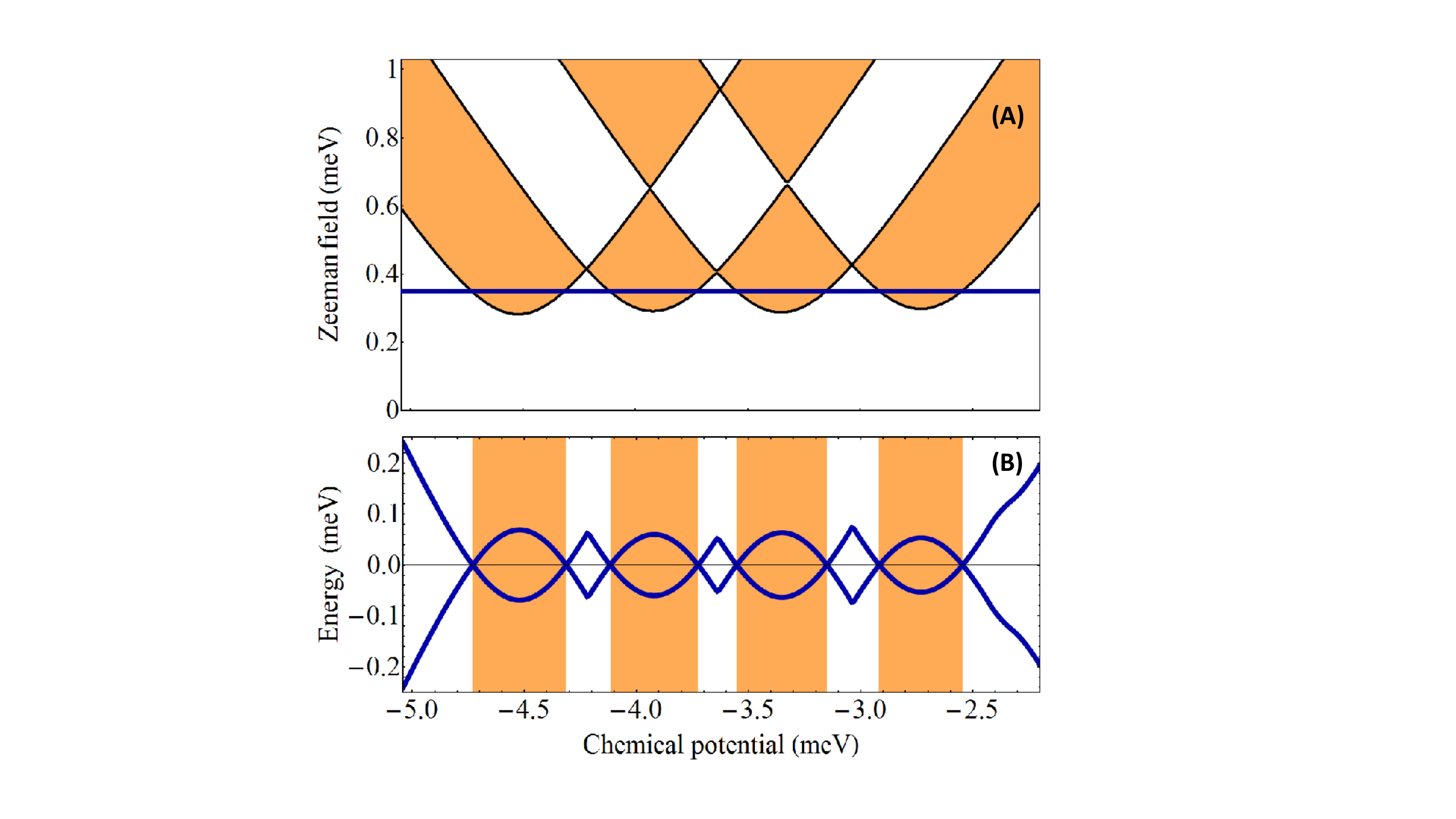}
\end{center}
\caption{
(A) Topological phase diagram for a square wire with $V_{\rm eff}(\ell)\neq 0$ and $\phi_\ell=0$. The white areas are topologically trivial and the orange regions are nontrivial. The values of the effective potential on the 8 chains are $ (0.5, 0, -0.5, -0.5, -0.5, 0, 0.5, 0.5)$ meV and the inter-chain hopping is $t^\prime=2.25~$meV.
(B) Evolution of the minimum quasiparticle gap along the horizontal cut $\Gamma=0.35~$ meV shown in the top panel.}
\label{Fig4TT}
\end{figure}

The existence of gapless superconducting phases in systems with rotation symmetry is generic, i.e. it holds for $n>3$. We emphasize that gapless phases cannot host stable Majorana modes and, therefore, they are not suitable for studying Majorana physics. Applying a symmetry-braking potential $V_{\rm eff}(\ell)\neq 0$ opens a finite gap throughout the entire phase diagram, except, of course, the phase boundaries, where the quasiparticle gap vanishes at $k=0$.  To better illustrate this point, we calculate the topological phase diagram for a square wire with $V_{\rm eff}(\ell)\neq 0$ and the minimum gap along a representative cut through the phase diagram. The results are shown in Fig. \ref{Fig4TT}. Note that all topologically trivial and nontrivial phases are gapped. However, the gaps are rather small indicating the fact that topological superconductivity (and the corresponding Majorana modes) are not very robust.  

An important difference between the phase diagram shown in Fig. \ref{Fig4TT} and that in Fig. \ref{Fig1TT} is that for the square wire we have used a larger value of the  inter-chain hopping, $t^\prime=2.25~$meV. Enhancing the coupling between chains widens the low-field  topological regions (which would practically vanish in the limit $t'/t \to 0$). Finally, we emphasize that although a finite system with parameters corresponding to a topologically nontrivial phase will support one pair of MZMs (i.e. one Majorana mode at each end of the wire), generically each Majorana mode is hosted by multiple chains (rather than a single chain). For example, in a configuration corresponding to Fig. \ref{Fig4TT}, the low-field topological phases with $\mu < 3.7~$meV can support MZMs hosted by chains 3 and 5 [with minimum values of $V_{\rm eff}(\ell)$], while for $\mu > 3.7~$meV the  MZMs are hosted by chains 1 and 7 [corresponding to the  maximum values of $V_{\rm eff}(\ell)$].


\section{Wires coupled with multiple superconductors: the stabilizing role of the phase difference}

\begin{figure} [h]
\begin{center}
\includegraphics[width=16.0cm]{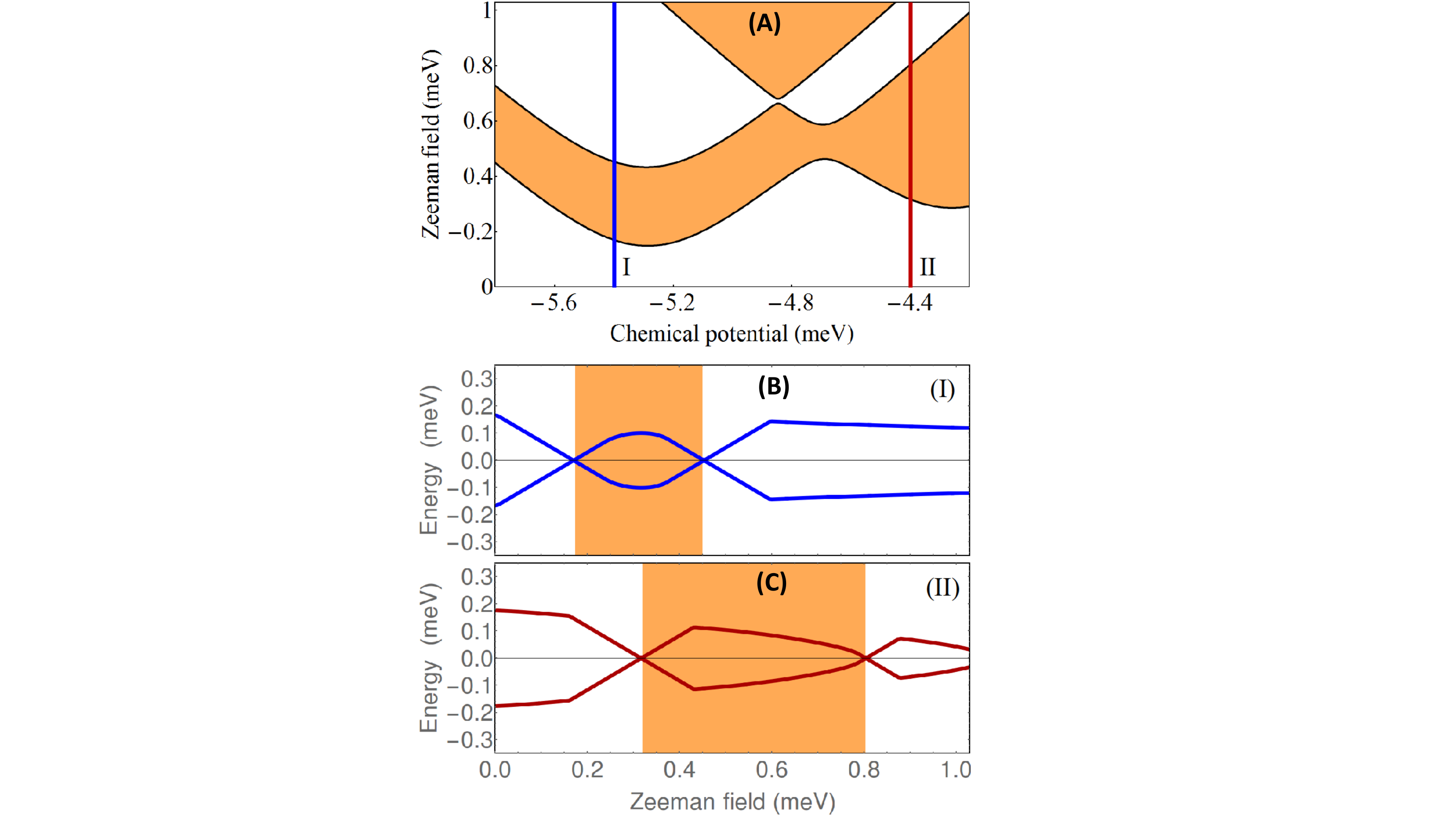}
\end{center}

\vspace{-5mm}
\caption{(A) Topological phase diagram for a triangular wire with $V_{\rm eff}(\ell)\neq 0$ and $\phi_1=0$, $\phi_3 = \pi/2$, $\phi_5=-\pi/2$. The white and orange phases are topologically trivial and nontrivial, respectively. The effective potential is the same as in Fig. \ref{Fig1TT}(B). (B) Dependence of the minimum quasiparticle gap on the Zeeman field along the blue cut (I)  in panel (A). (C) Dependence of the minimum quasiparticle gap on the Zeeman field along the dark red cut (II) in panel (A).  Note the increased  stability of the low-field topological phase [see for comparison Fig. \ref{Fig1TT}(B)] and the fact that the minimum critical field $\Gamma_B^c\approx 0.15~$meV is lower than the pairing potential for corner chains, $\Delta=0.3~$meV.}
\label{Fig5TT}
\end{figure}

A critical question that we want to investigate concerns the effect of a nonzero superconducting phase difference in a wire coupled to multiple parent superconductors. A non-zero phase difference was shown to stabilize the topological phase in a Josephson junction across a 2D electron gas with Rashba spin-orbit coupling and in-plane magnetic field \cite{Pientka2017} and in a topological insulator nanoribbon coupled with two superconductors \cite{Sitthison2014}. Here, for concreteness, we consider a triangular core-shell nanowire modeled by six chains, as described above, which are coupled to three separate superconductors that induce pairing potentials characterized by $\phi_1=0$, $\phi_3 = \pi/2$, and $\phi_5=-\pi/2$. The other parameters are the same as in  Fig. \ref{Fig1TT}(B), i.e. the case $V_{\rm eff}\neq 0$ discussed above.
The corresponding phase diagram is shown in Fig. \ref{Fig5TT}. Remarkably, the ``crossing points'' that characterize the phase diagram in Fig. \ref{Fig1TT} disappear  and, upon increasing the Zeeman field,  we have an alternance of trivial and nontrivial phases for all values of the chemical
potential. More importantly, the low-field topological phase becomes stable for a wide range of chemical potentials, i.e., it is characterized
by a significant quasiparticle gap, as shown in panels (B) and (C). In addition, the lowest critical field $\Gamma_B^c\approx 0.15~$meV is
about half the value of the pairing potential (i.e. $\Delta/2$). This is in sharp contrast with the case of hybrid systems involving a single
superconductor, or multiple superconductors having the same phase, $\phi_\ell={\rm const.}$, where the minimum critical field is  $\Gamma_B^c =
\Delta$. 

\begin{figure}
\begin{center}
\includegraphics[width=14.0cm]{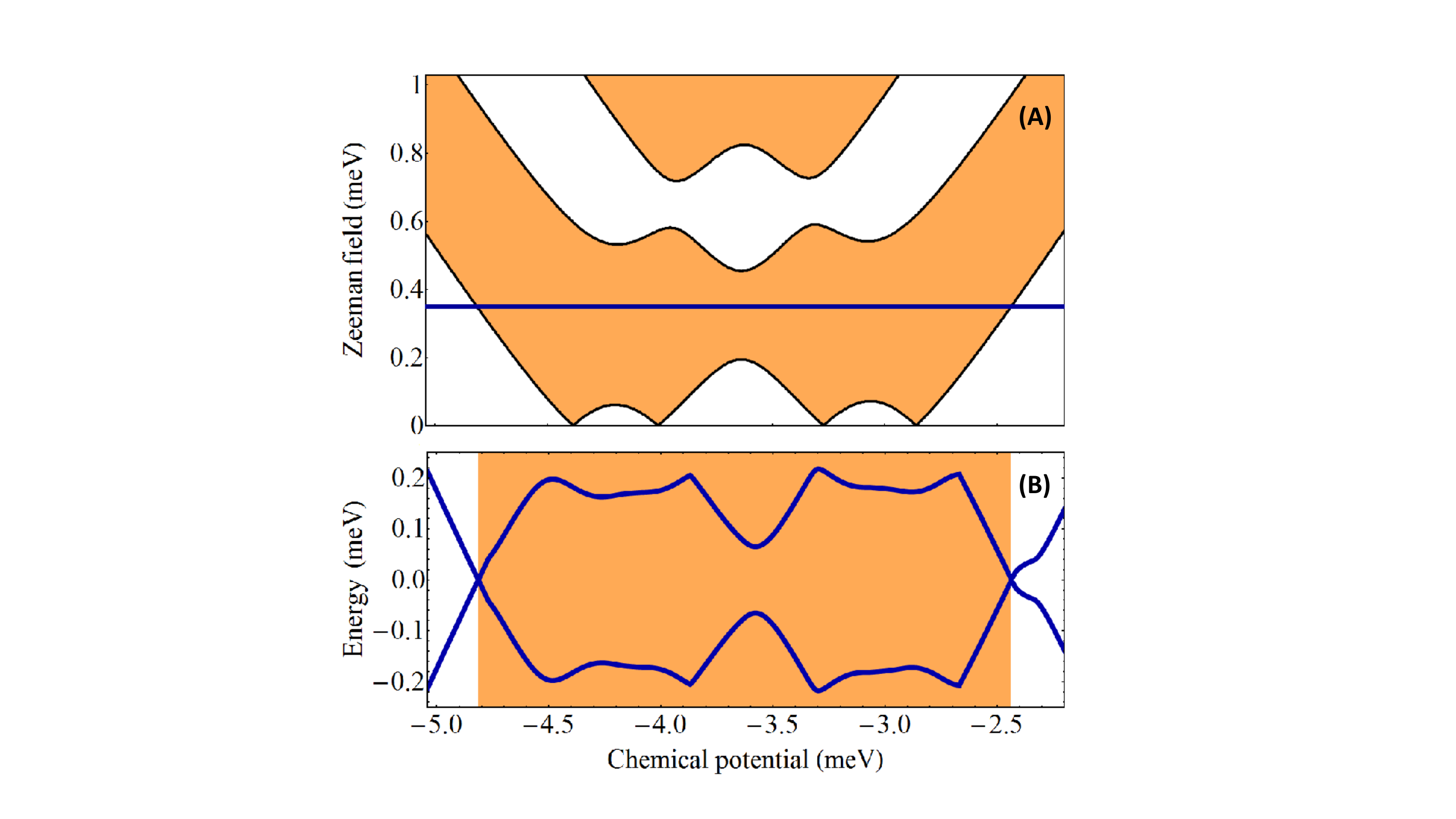}
\end{center}

\vspace{-5mm}
\caption{
(A) Topological phase diagram for a square wire with $V_{\rm eff}(\ell)\neq 0$ and $\phi_1=\pi/2$, $\phi_3 = -\pi/2$, $\phi_5=\pi/2$, and $\phi_7 = -\pi/2$. The white areas are topologically trivial and the orange regions are nontrivial. The values of $V_{\rm eff}(\ell)$ and the inter-chain hopping $t^\prime$ are the same as in Fig. \ref{Fig4TT}. 
(B) Evolution of the minimum quasiparticle gap along the horizontal cut $\Gamma=0.35~$ meV shown in the top panel. Note the significant expansion of the low-field topological phase [see for comparison Fig. \ref{Fig4TT}], the large topological gap, and the low values of the critical field.}
\label{Fig6TT}
\end{figure}

A comparison between the results in Fig. \ref{Fig1TT} and those in Fig. \ref{Fig5TT} suggests that the superconducting phase could be used as a  knob for tuning the system across a topological quantum phase transition. For example, if $\mu=-5.4~$meV and $\Gamma_B=0.25~$meV the
system evolves as a function of the superconducting phase differences from a topologically-trivial state when $\phi_\ell=0$ to a topological
superconductor when $\phi_1=0$ and $\phi_3=-\phi_5=\pi/2$. We emphasize that the simplified tight-binding model can only provide a qualitative
picture of the low-energy physics of proximitized core-shell wires. For quantitative predictions regarding the dependence of the low-energy
physics on the effective bias potential $V_{\rm eff}$ and the superconducting phases $\phi_\ell$ a more detailed modeling of the hybrid structure
(possibly, at the microscopic level) is necessary.	

To corroborate our findings regarding the effect of a phase difference, we consider the square wire corresponding to the phase diagram shown in Fig. \ref{Fig4TT} coupled to four separate superconductors that induce pairing potentials characterized by $\phi_1=\pi/2$, $\phi_3 = -\pi/2$, $\phi_5=\pi/2$, and $\phi_7 = -\pi/2$. The corresponding phase diagram is shown in Fig. \ref{Fig6TT}. The qualitative effect of having finite phase differences is the same as in the case of the triangular wire, while quantitatively it is more significant as a results of a stronger inter-chain coupling $t^\prime$. The topology of the phase diagram is similar to that shown in Fig. \ref{Fig5TT}. However, the low-field topological phase now occupies a significant region of the parameter space and the minimum critical field $\Gamma_B^c$ is practically zero. Furthermore, the topological gap is substantial, as shown in the lower panel of Fig. \ref{Fig6TT}, indicating a robust topological superconducting phase.

\newpage

\section{Majorana modes in finite core-shell nanowires}

\begin{figure}
\begin{center}
\includegraphics[width=8.0cm]{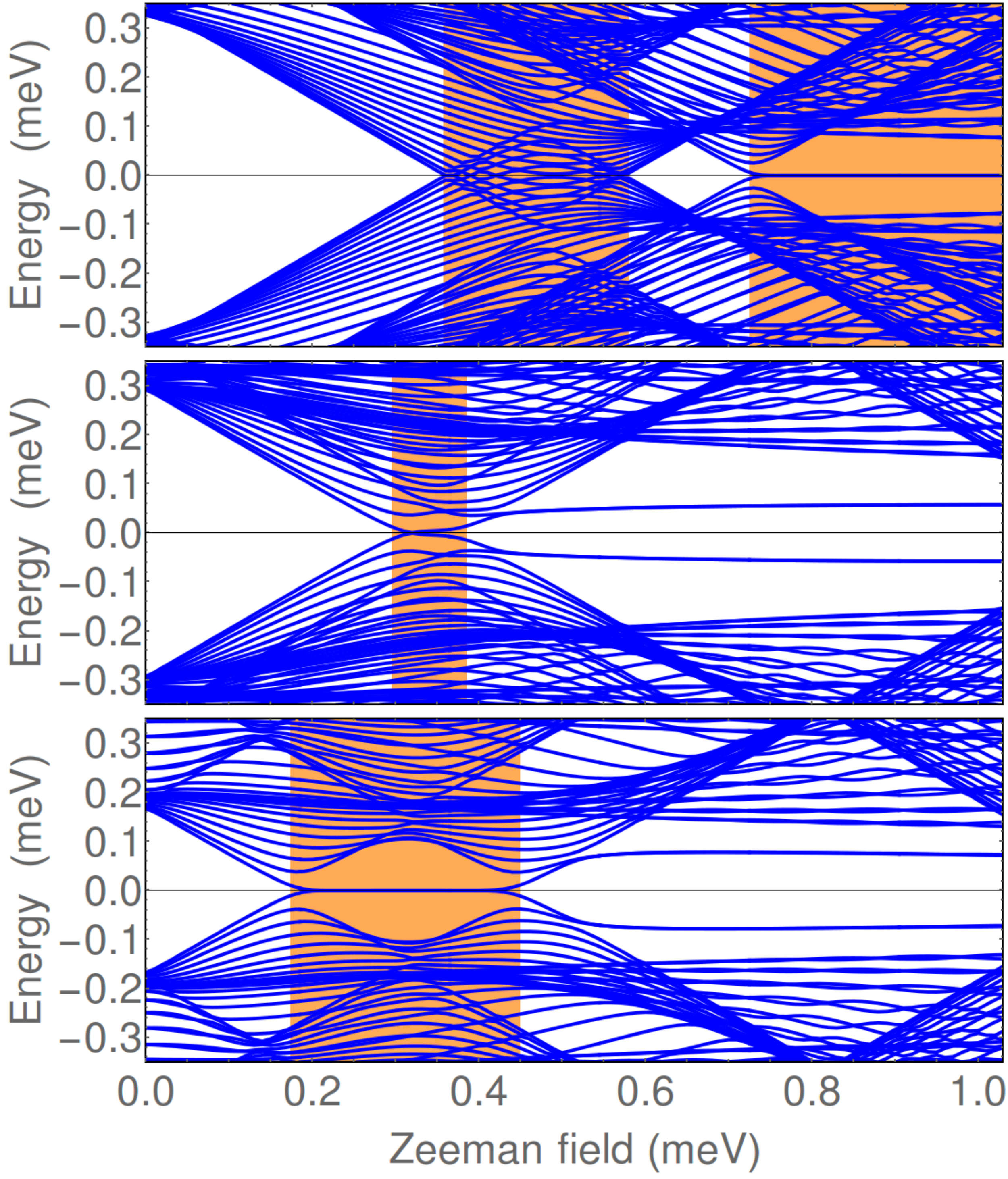}
\end{center}
\caption{Dependence of the low-energy spectrum on the Zeeman field
for  a finite triangular  wire of length $L=2.25~\mu$m and  chemical
potential $\mu=-5.4~$meV.  The parameters used in the top, middle,
and bottom panels correspond to Fig. \ref{Fig2TT}(I-A),
Fig. \ref{Fig2TT}(I-B), and Fig. \ref{Fig5TT}(B), respectively.}
\label{Fig7TT}
\end{figure}

As a consistency check for the results discussed above, which are based
on a translation-invariant model (i.e., infinite wire), and to gain
further insight into the low-energy physics of the hybrid structure,
we continue now with the case of wires of finite length.   For concreteness, we
consider a triangular wire of \blue{length $L=2.25~\mu$m} in the parameter
regimes corresponding to the panels labeled by ``I'' and ``II''' in
Figs. \ref{Fig2TT}, \ref{Fig3TT}, and \ref{Fig5TT}.
The dependence of
the  low-energy spectrum on the Zeeman field for $\mu=-5.4~$meV, i.e.,
corresponding to the (I) panels, is shown in Fig. \ref{Fig7TT}. Note that
when $V_{\rm eff}=0$ and $\phi_\ell=0$ (top panel) the first transition is
from a topologically-trivial phase to a gapless superconductor, as already
discussed in the context of Fig. \ref{Fig2TT}. The high-field topological
phase ($\Gamma_B>0.7~$meV) is characterized by a zero-energy Majorana
mode separated by a finite gap from finite energy excitations. Applying
a symmetry-breaking potential $V_{\rm eff}$ (middle panel) generates
a low-field topological phase characterized by a small bulk gap and
a weakly stable, energy-split Majorana mode. However, the stability
of this topological phase can be significantly enhanced by creating
phase differences between the parent superconductors (bottom panel).
Note that in the middle and bottom panels the second trivial phase
($\Gamma_B$ larger than about $0.35~$meV and $0.45~$meV, respectively)
is characterized by sub-gap states that can be viewed as pairs of
overlapping, energy split Majorana bound states (at each end of the
wire). This result suggests that coupling the nanowire to multiple
parent  superconductors and controlling their relative phases represents
a powerful scheme for enhancing the robustness of the topological phase
and tuning the system across a topological quantum phase transition.

\begin{figure}
\begin{center}
\includegraphics[width=8.0cm]{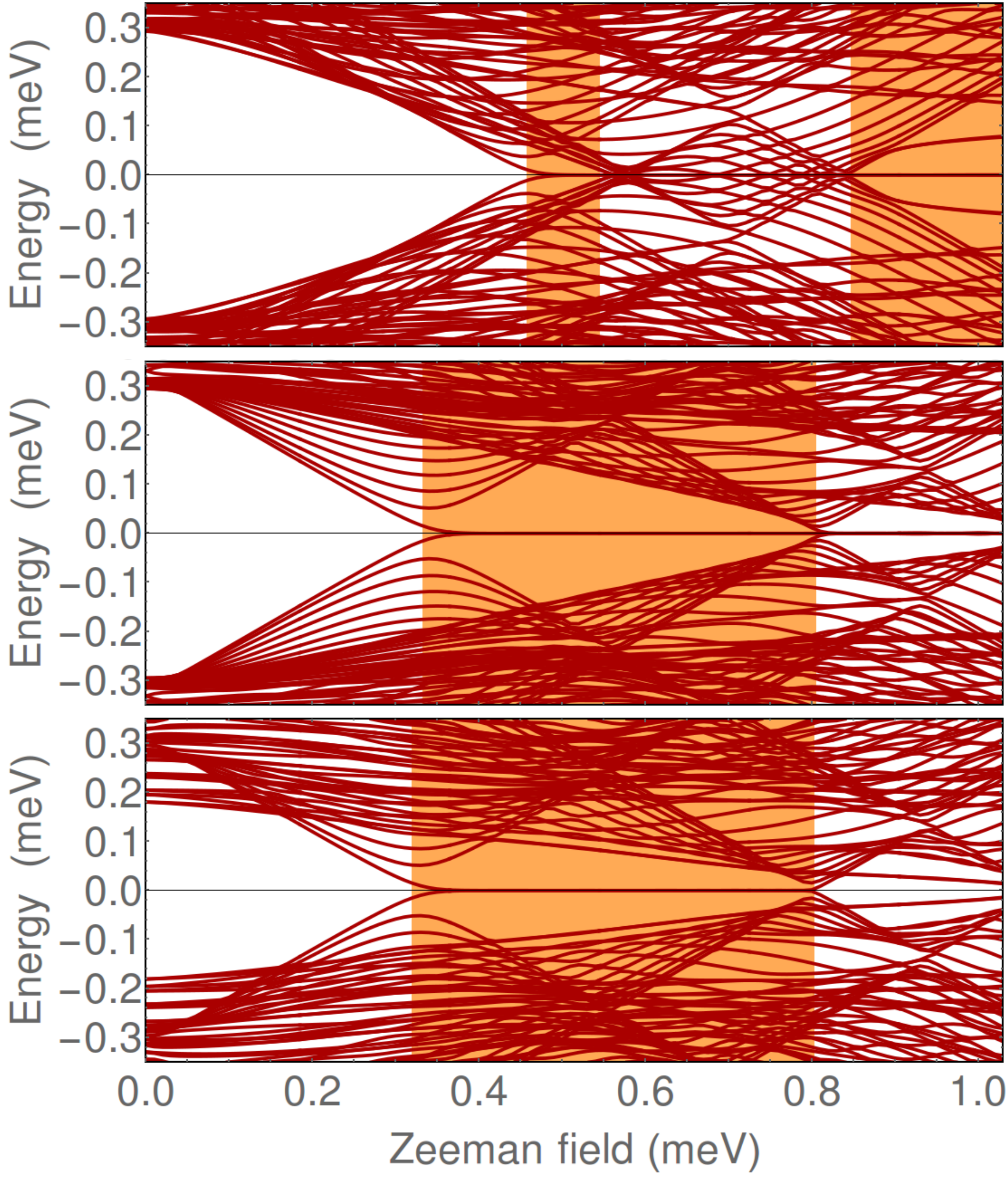}
\end{center}
\caption{Dependence of the low-energy spectrum on the Zeeman field for
a finite triangular wire of length $L=2.25~\mu$m and chemical potential
$\mu=-4.4~$meV. The parameters used in the top, middle, and bottom panels
correspond to Fig. \ref{Fig3TT}(II-A), Fig. \ref{Fig3TT}(II-B), 
and Fig. \ref{Fig5TT}(C), respectively.}
\label{Fig8TT}
\end{figure}

%
The low-energy spectra for $\mu=-4.4~$meV, i.e. those corresponding to
the (II) panels in Figs. \ref{Fig3TT} and \ref{Fig5TT},  are shown in
Fig. \ref{Fig8TT}.  In the top panel, note the presence of a gapless
superconducting phase, which is consistent with our conclusions based on
the results shown in Fig. \ref{Fig3TT}. Also note that the high-field
topological phase ($\Gamma_B > 0.85~$meV) supports two finite energy
sub-gap modes, in addition to the zero-energy Majorana mode. Again, we
can interpret these modes as pairs of overlapping Majoranas. We conclude
that in this phase the hybrid system has three Majorana bound states at
each end of the wire, two Majorana modes acquiring finite energy and
one remaining gapless, consistent with a ${\mathbb Z}_2$ topological
classification. Applying a symmetry-breaking potential (middle panel)
enhances significantly the stability of the low-field topological phase
and generates a second trivial phase ($\Gamma_B>0.9~$meV) that is gapped
in the bulk, consistent with Fig. \ref{Fig3TT}. Remarkably, this trivial
phase supports a pair of zero-energy Majorana modes at each end of
the wire, which correspond to the mid-gap states visible in the middle
panel of Fig. \ref{Fig8TT}. This indicates the presence of an additional
``hidden'' symmetry in the system, which makes it an element of the BDI
symmetry class \cite{Dumitrescu2015}.  This symmetry is broken in the
presence of a superconducting phase difference (bottom panel), when the
sub-gap modes acquire finite energy.
%
%

\newpage

\section{Symmetry and gapless superconducting phases} 

\blue{
The existence of the gapless superconducting phases (indicated by the star in the top
panels of Fig. \ref{Fig1TT} and Fig. \ref{Fig2TT}) is a consequence of the threefold rotation
symmetry of the triangular wire with $V_{\rm eff}(\ell)=0$ and identical
superconductors.  Breaking this symmetry automatically opens a (bulk)
gap in the spectrum.  To illustrate this property we consider the system
of finite length $L=2.25$ nm, with the other parameters corresponding to 
Fig. \ref{Fig1TT}(A), with chemical potential $\mu=-5.4~$meV
(i.e. the blue vertical line there), and $V_{\rm eff}(\ell)=0$,
and we focus on the gapless phase $0.36 <\Gamma_B <0.58~$meV. The low-energy
spectrum is shown in Fig. \ref{Fig1exTT}(A), which is in fact a zoom into
the top panel of Fig. \ref{Fig7TT}. 
We consider now a small symmetry-breaking potential, with the same proportions
as in Fig. \ref{Fig1TT}(B), Fig. \ref{Fig2TT}(I-B), and middle panel of Fig. \ref{Fig7TT},
but now ten times weaker, i.e.  $V_{\rm eff} =V_0(2, 0.5, -1, -1, -1, 0.5)$ with $V_0=33.3~\mu$eV.
The potential opens a bulk gap that hosts a mid-gap Majorana mode, as shown in
Fig. \ref{Fig1exTT}(B).  To emphasize that the opening of a bulk gap is
the result of breaking the threefold rotation symmetry, we also consider
a system with vanishing effective potential,  $V_{\rm eff}(\ell)=0$, in which
we break the symmetry by coupling the wire to parent superconductors
having different bulk gaps, so that the proximity-induced pairing
potentials for the edges are $\Delta_1=0.375~$meV,  $\Delta_3=0.300~$meV,
and $\Delta_5=0.300~$meV. Here we do not consider any relative phase 
between the superconductors.  Again, a small bulk gap opens in the (bulk)
spectrum and a (nearly-zero) Majorana mode emerges as a mid-gap state,
as can be seen Fig. \ref{Fig1exTT}(C).  }

\blue{
\begin{figure}
\begin{center}
\includegraphics[width=12.0cm]{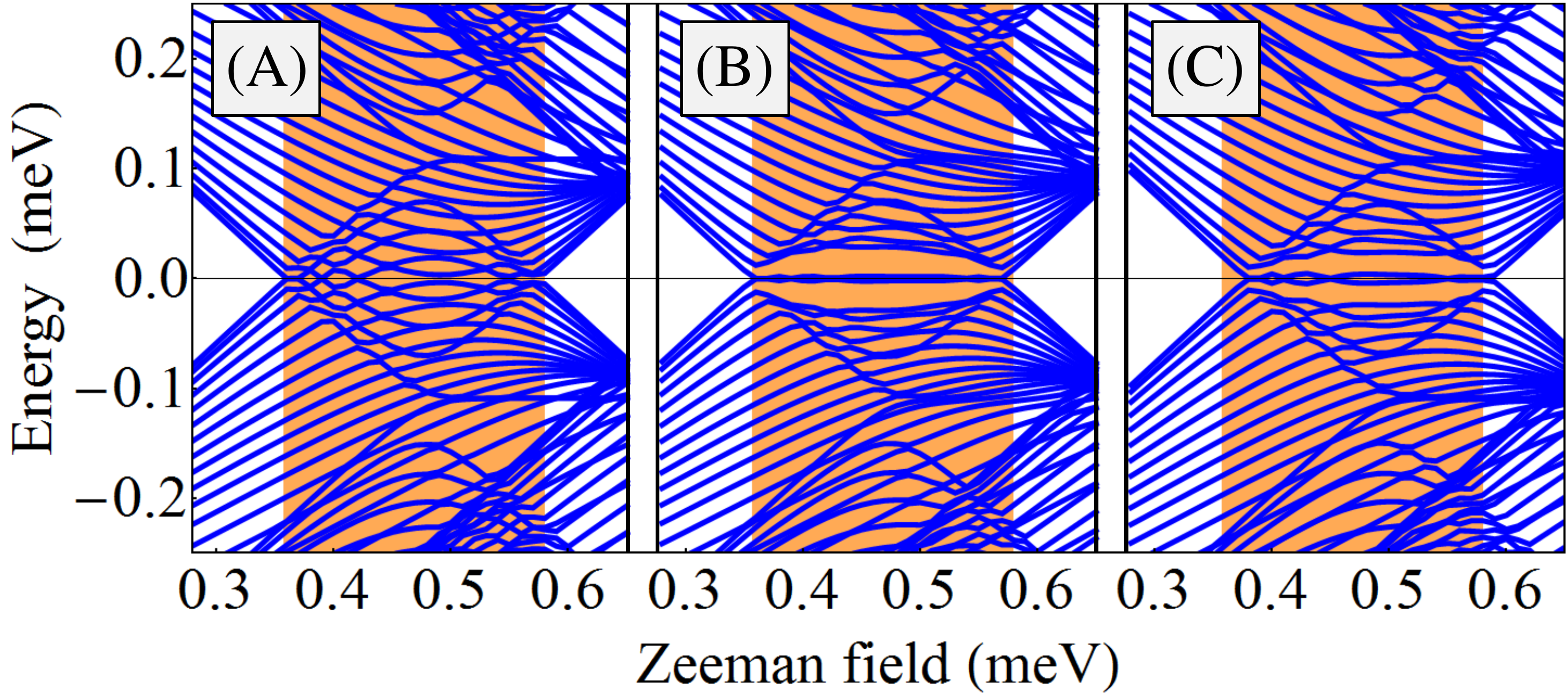}
\end{center}
\caption{Low-energy spectra as a function of the Zeeman field for
a finite triangular wire of length  $L = 2.25~\mu$m and chemical
potential $\mu=-5.4~$meV. (A) Gapless superconducting phase in a
system with threefold rotation symmetry, like in Figs. \ref{Fig1TT}(A) and \ref{Fig2TT}(I-A). 
(B) Applying a symmetry-breaking $V_{\rm eff}$, ten times waker than in Fig. \ref{Fig1TT}(B),
a small  bulk gap develops, like in Fig. \ref{Fig2TT}(I-B), that hosts a mid-gap Majorana mode. 
(C) Symmetry broken by coupling
the wire to different superconductors inducing edge pairing potentials
$\Delta_1=0.375~$meV,  $\Delta_3=0.3~$meV, and $\Delta_5=0.3~$meV. The
filled (orange) region $0.36 <\Gamma_B <0.58~$meV represents the
topological superconducting phase (of an infinite wire) in the presence
of an infinitesimally-small symmetry-breaking perturbation.}
\label{Fig1exTT}
\end{figure}
}

\blue{
Another important general property of the Majorana modes illustrated
in Fig. \ref{Fig1exTT}, panels (B) and (C), is the presence of energy
splitting oscillations \cite{Cheng2009,DSarma2012}. In general, the energy splitting
is caused by a finite overlap of the Majorana modes localized at the
opposite ends of the wire. The amplitude of the oscillations depends
on the Majorana localization length $\xi$ \cite{DSarma2012}, which increases as
the topological gap decreases,  diverging in the gapless limit. This
behavior is illustrated in Fig. \ref{Fig2exTT}. The top panel represents
the lowest-energy state corresponding to a gapless system with threefold
rotation symmetry (i.e., $V_{\rm eff}=0$), which could be seen as a linear
combination of Majorana modes with an infinite characteristic lenghscale,
$\xi\rightarrow \infty$. Introducing a symmetry-breaking perturbation
($V_{\rm eff}\neq 0$) opens a (bulk) topological gap that increases with
increasing the effective potential. In addition, in a finite system a
midgap state emerges, consisting of two (partially) overlapping Majorana
modes localized at the opposite ends of the wire.  As clearly shown
in Fig. \ref{Fig2exTT}, the characteristic length scale $\xi$ of the
Majorana modes decreases as the amplitude $V_0$ of the symmetry-breaking
potential increases (i.e., as the topological gap increases).
}

\blue{
\begin{figure}
\begin{center}
\includegraphics[width=7.0cm]{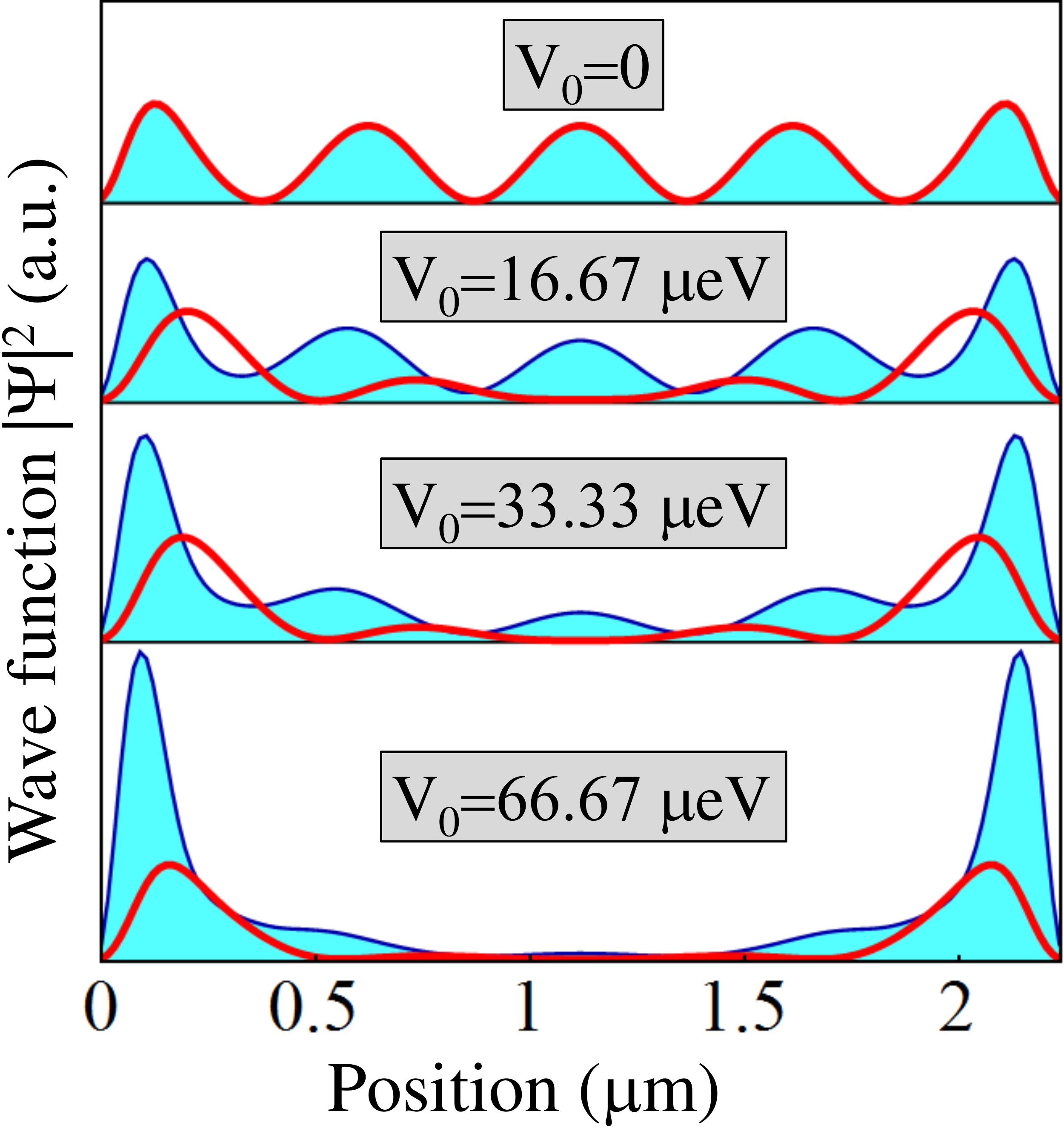}
\end{center}
\caption{Position dependence of the lowest energy wave function
corresponding to a finite triangular wire of length  $L = 2.25~\mu$m,
chemical potential $\mu=-5.4~$meV, Zeeman field $\Gamma_B=0.45~$meV,
and symmetry-breaking effective potential with amplitude $V_0$
[see Fig. \ref{Fig1exTT}(B)]. The thick (red) line represents the
probability distribution $|\Psi_1(x)|^2$ along the edge $\ell=1$, while
the filled (blue) line represents the probability distribution along the edges
$\ell=3,5$. With increasing the amplitude of the symmetry-breaking potential
the (bulk) topological gap increases, which leads to the reduction of the characteristic
length $\xi$ of the Majorana modes localized at the opposite ends of
the wire.}
\label{Fig2exTT}
\end{figure}
}

\blue{
We note that, from the perspective of quantum computation, the zero-energy
Majorana modes have to be i) well separated spatially (to minimize the
overlap and, consequently, the energy splitting $\delta E$) and ii)  well
separated in energy from all other low-energy states (by a certain minimum
quasiparticle gap $\Delta E$). The first condition ensures that the
Majorana modes have non-Abelian properties, while the second guarantees
that the parity of the low-energy Majorana sub-space is fixed
(the presence of other  low-energy states would allow excitations from
the Majorana sub-space, which would change its parity and destroy any
quantum information stored in the Majorana system).  If these conditions
are satisfied, the Majorana modes span a nearly-zero energy subspace
that can be used for storing and processing quantum information. The
characteristic timescale $\tau$ for quantum operations has to satisfy
the condition $\delta E \ll \hbar/\tau \ll \Delta E$. Of course,
the impossibility of satisfying this condition is manifest in regimes
characterized by small topological gaps, as $\delta E$ and $\Delta E$
become comparable in the gapless superconductor limit.
}

\section{Effects of disorder}

\begin{figure} 
\begin{center}
\includegraphics[width=7.5cm]{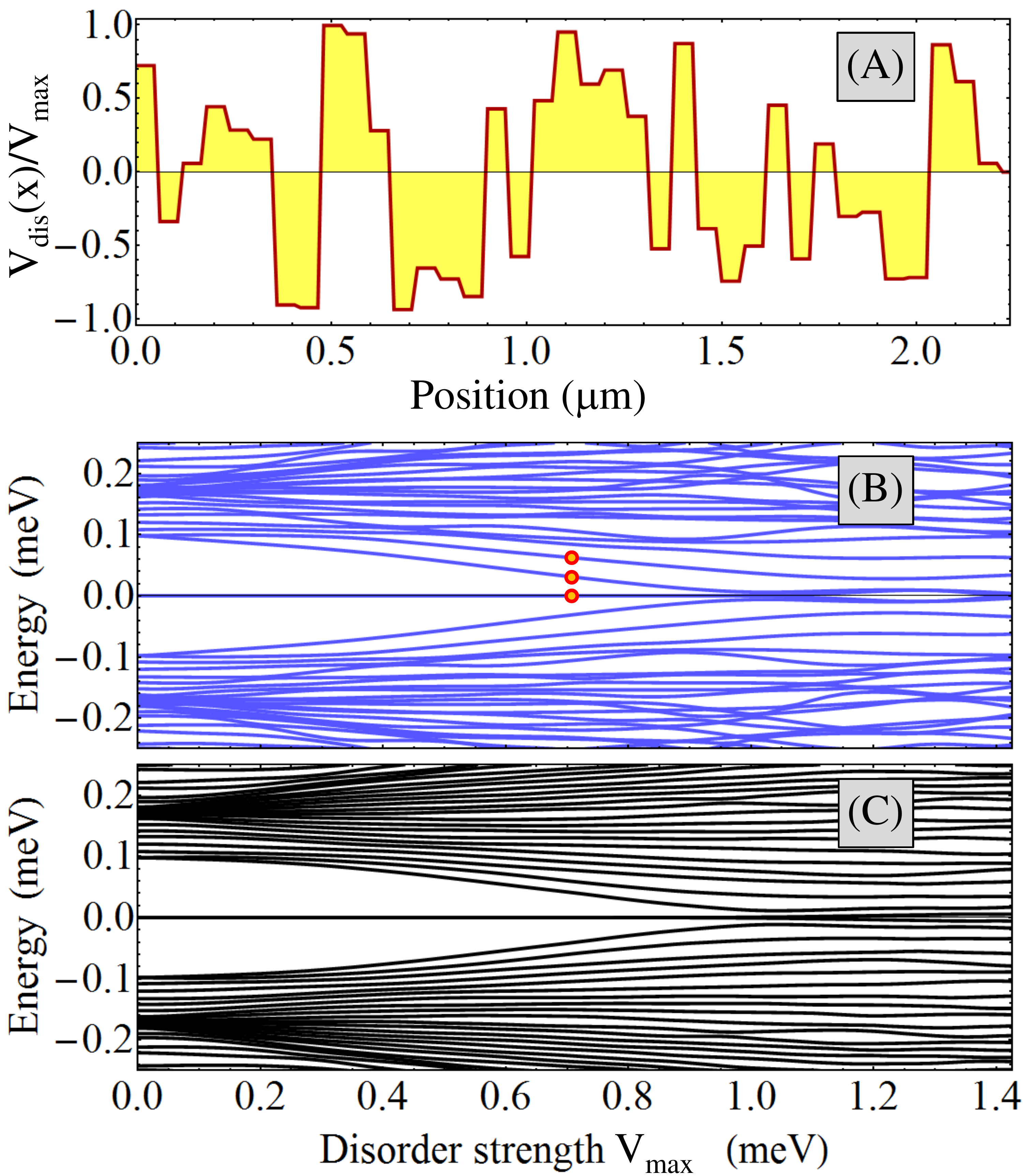}
\end{center}
\caption{(A) 
\blue{Position dependence of the normalized disorder potential
along the edge $\ell=3$ of a triangular wire for a specific disorder
realization. The disorder profiles along the edges $\ell=1,5$ (not shown)
are different, but characterized by similar qualitative features. In
particular, the characteristic length scale for the  potential variations
is $\delta_d=60~$nm. (B) Dependence of the low-energy spectrum on the
amplitude $V_{\rm max}$ of the disorder potential for the disorder
realization shown in panel (A). (C) Low-energy spectrum averaged over
$50$ different disorder realizations as a function of $V_{\rm max}$. The
parameters of the system are: 
wire length  $L = 2.25~\mu$m, chemical potential $\mu=-5.4~$meV,
effective potential $V_{\rm eff} =(0.67, 0.17, -0.33, -0.33, -0.33,
0.17)~$meV, superconducting phases $\phi_1=0$, $\phi_3=\pi/2$,
$\phi_5=-\pi/2$ and Zeeman field $\Gamma_B=0.35~$meV.}
}
\label{Fig3exTT}
\end{figure}

\blue{
Another element that can compromise the topological protection of
the Majorana subspace is the presence of disorder. Generically,
disorder induces low-energy sub-gap states, thus reducing $\Delta E$
\cite{Bagrets12,Liu12,DeGottardi13,Rainis13,Adagideli14}. 
The effect of potential disorder on a topological phase
realized in a triangular wire is illustrated in Fig. \ref{Fig3exTT}.
Panel (A) shows the position dependence (along the wire) of a typical
disorder potential $V_{\rm dis}(x)$ considered in the calculation. Next,
we calculate the low-energy spectrum in the presence of a disorder
potential with a fixed profile but a varying amplitude $V_{\rm max}$
[see  Fig. \ref{Fig3exTT}(B)]. As the disorder strength increases,
several low-energy states converge toward zero-energy, so that the
quasiparticle gap $\Delta E$ practically collapses when the amplitude
of the effective disorder potential is larger than $V_{\rm max} \approx
1~$meV. To demonstrate that this is not an accidental property of a
specific disorder realization, we also calculated the spectrum averaged
over multiple disorder realizations [see  Fig. \ref{Fig3exTT}(C)]. The
qualitative features discussed above are manifestly present. We note
that ``critical'' disorder strength associated with the collapse of
the quasiparticle gap depends on the characteristic length scale of the
disorder potential, as well as the topological gap of the clean system,
larger gaps implying an increased robustness against disorder.
}

\begin{figure}
\begin{center}
\includegraphics[width=5.5cm]{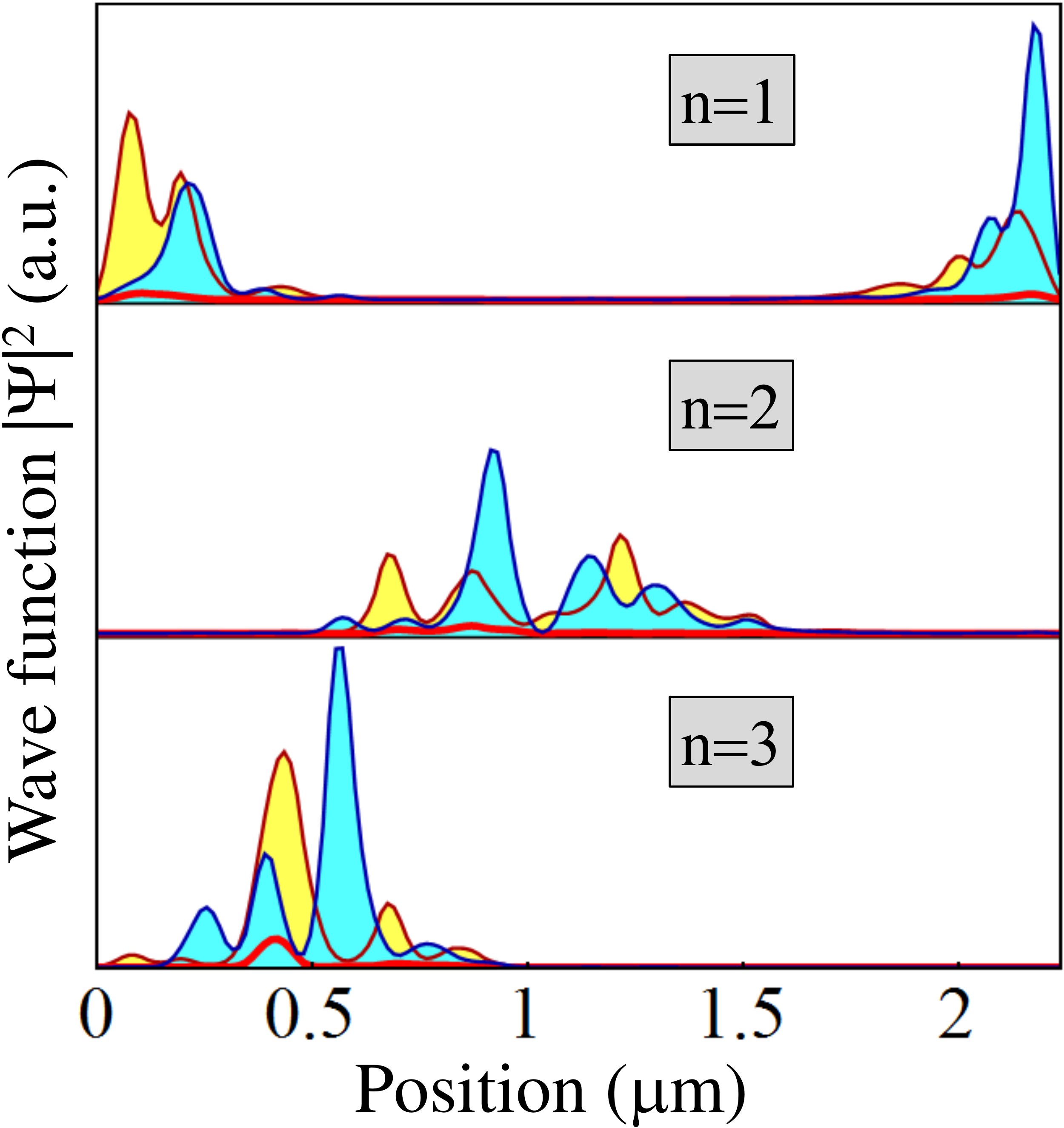}
\end{center}
\caption{Spatial profiles of the three lowest energy states corresponding
to the red dots in Fig. \ref{Fig3exTT}(B). The thick (red) line represents
the profile along the edge $\ell=1$, while the filled lines represent the
profiles along the edge $\ell=3$ (blue/ light blue filling) and $\ell=5$
(dark red/ yellow filling).}
\label{Fig4exTT}
\end{figure}

The final point that we want to address concerns the structure of
the disorder-induced low-energy states. Specifically, we calculate the
spatial profiles of the three lowest-energy states marked by red dots in
Fig. \ref{Fig3exTT}(B). The results are shown in Fig. \ref{Fig4exTT}. We
note that the Majorana modes ($n=1$) are well localized near the opposite
ends of the wire and have most of the spectral weight on the edges
$\ell=3,5$ as a result of applying a bias potential $V_{\rm eff}(\ell)$. The
disorder-induced states ($n=2,3$) are localized inside the wire and have
most of their spectral weight on the same edges, $\ell=3,5$. We conclude
that the presence of disorder induces low-energy localized states than
can destroy the topological protection of the Majorana subspace. We note
that within a topological quantum computation scheme based on qubits
characterized by a finite charging energy \cite{Aasen16,Karzig17}, interaction-mediated
transitions between the Majorana modes and disorder-induced localized
states are possible even when the spatial overlap of the two types of
states is exponentially small. Such transitions, which create low-energy
quasiparticles,  could completely compromise the topological protection
of the quantum computation scheme.

\newpage

\section{Geometrical model of a prismatic shell} 

\begin{figure} [h]
\begin{center}
\includegraphics[width=10.0cm]{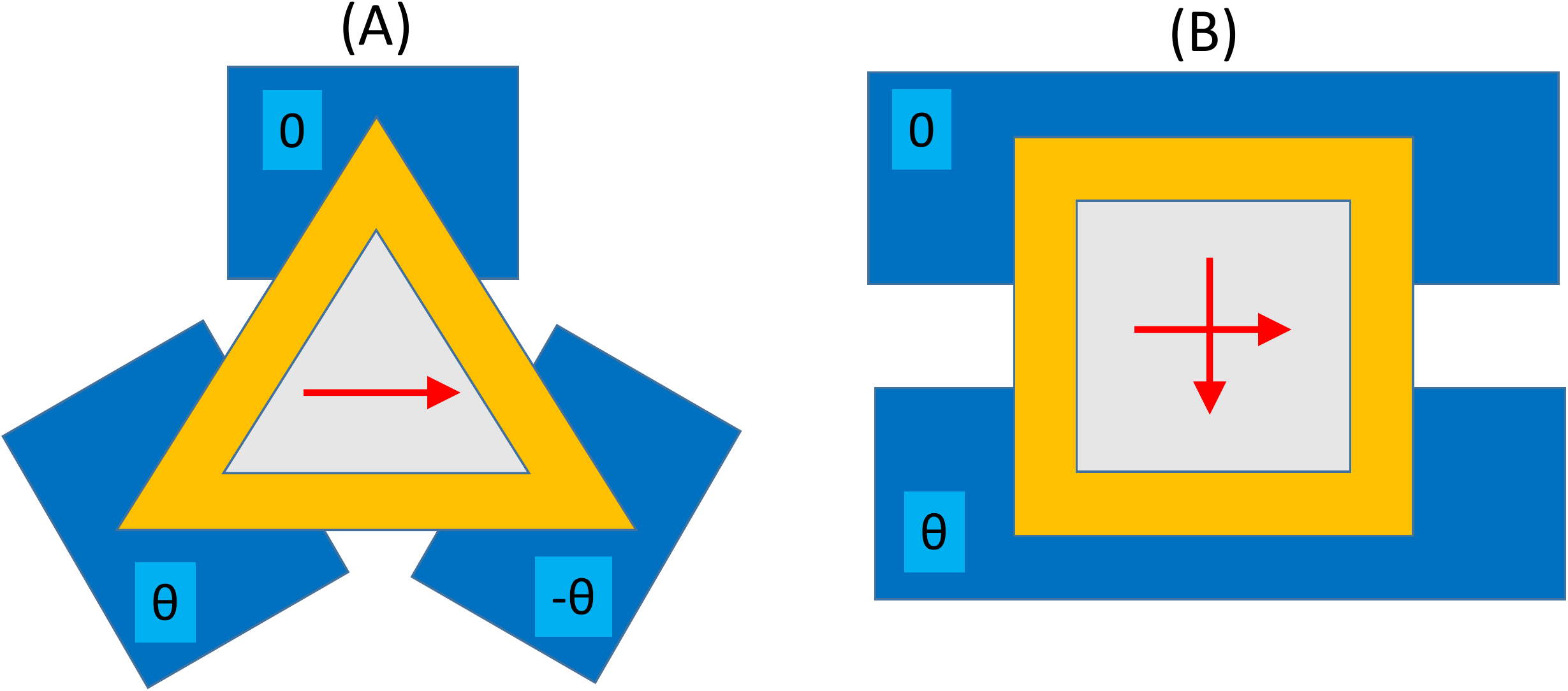}
\end{center}
\caption{A schematic cross section of the hybrid
semiconductor-superconductor experimental device incorporating a
core-shell wire. The core is shown in grey and the shell in yellow.
The blue blocks represent the superconductor metals attached to the wire.
The lower superconductors can have phases $\pm \theta$ relatively to
the upper one considered with zero phase.  The red arrows indicate the
electric field included in our geometrical model.  (A) In the triangular
case it is parallel to one side of the triangle.  (B) In the square case
it can be either perpendicular or parallel to the superconductors.}
\label{Sample}
\end{figure}

In this section we analyze the results of a finer-grained model of
triangular and square prismatic shells, based on a geometrical description
\cite{Manolescu17}. First the two-dimensional Hamiltonian of a single
electron confined on the polygonal cross section is discretized on a grid
defined in polar coordinates and diagonalized numerically \cite{Sitek15,Sitek16}.  
The resulting low-energy eigenstates, corresponding to corner localization, 
are further used as a basis to find the eigenstates of the BdG Hamiltonian,
assuming plane waves in direction longitudinal to the prism.  The basis
includes the spin and the isospin.  The variable Zeeman energy is generated
by a uniform magnetic field $B$ longitudinal to the wire.  In addition we consider 
a relatively weak electric field $E$ transverse to the wire as a technical tool to break the
symmetry of the polygon, indicated by the red arrows in Fig.\ \ref{Sample}. 
This field is equivalent with the chain dependent potential $V_{\rm eff}(\ell)$ introduced
before.  First, a perfectly symmetric shell is experimentally unrealistic
from fabrication.  Second, as already mentioned, in a regular experimental
setup external gates and other contacts may affect the wire symmetries.
Third, a generic electric field can be seen as a tunable parameter that can
change the topological phase diagram.

We characterize the lateral size of the wire with the radius $R$ of a
circle surrounding the shell, and with the shell thickness $d$.  In the
present calculations we use $R=50$ nm for both geometries, but $d=12.5$ nm 
for the triangular shell and $d=8$ nm for the square shell.  
\blue{These values are comparable to the dimensions of the realistic 
core-shell nanowires mentioned in the experimental papers 
\cite{Wong11,Blomers13,Qian12,Heurlin15,Yuan15}.}
The material
parameters of the shell are chosen as for InSb.  For these geometric
parameters and with $m_{\rm eff}=0.014$ the energy separation between
the corner and side states is about 41 and 38 meV for the triangular and 
square case, respectively, meaning that
for these parameters the low energy physics can be very well
described by the corner states.   Therefore we can use a Rashba SOC model
similar to that of the planar electron gas, but on a cylindrical surface
of radius $R$, i.e., transformed from Cartesian to polar coordinates
\cite{Bringer11}.  Since the sides of the triangular shell are unpopulated
this model is qualitatively reasonable, and can lead to Majorana states.
As mentioned before a more elaborated microscopic description of the SOC
is beyond the scope of the present paper, and here we
simply adopt in the numerical calculations the coupling constant 
of bulk InSb, of 50 meV/nm.

\begin{figure}
\begin{center}
\includegraphics[width=6.0cm]{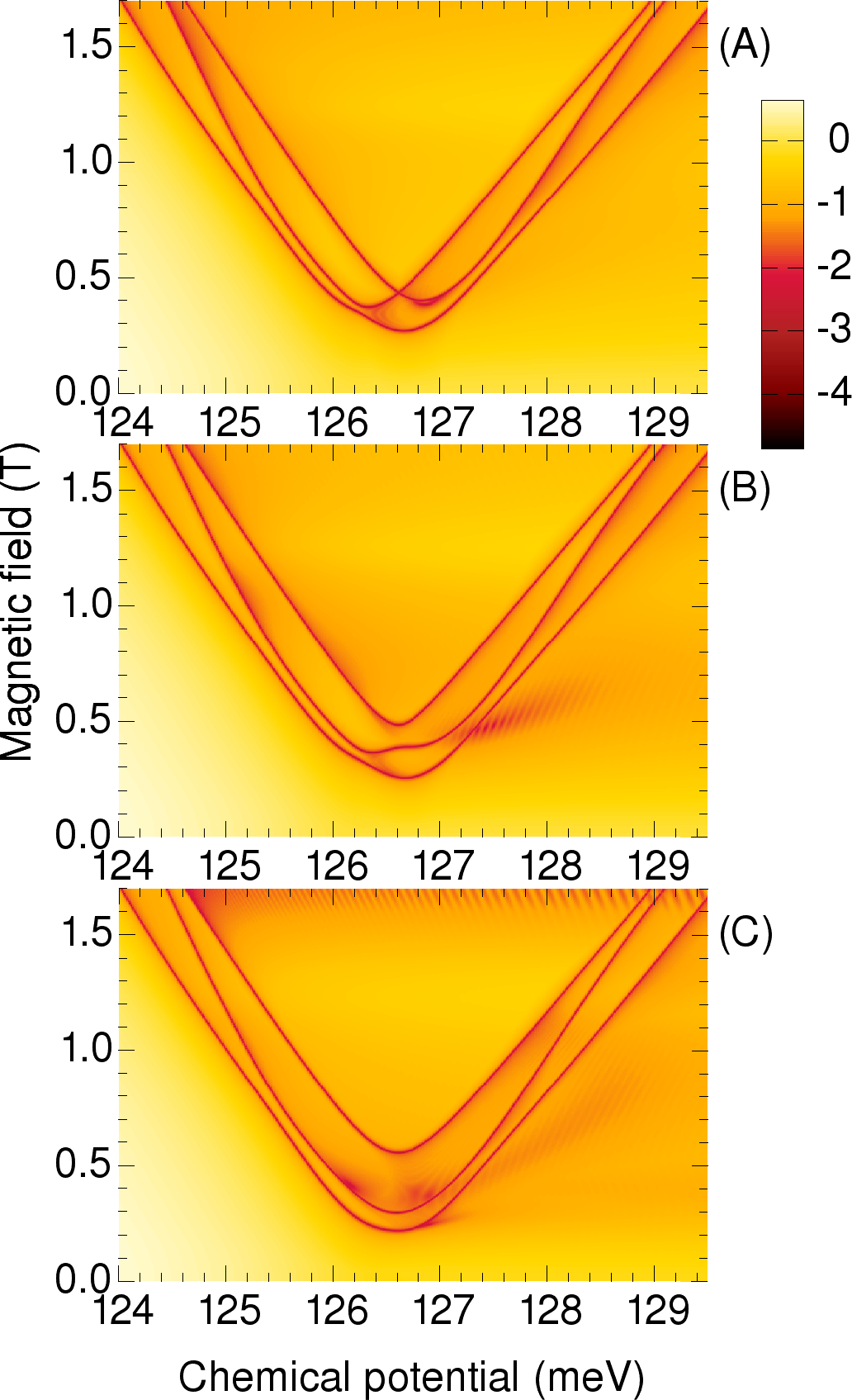}
\end{center}
\caption{Phase \blue{boundaries} for the triangular wire in the corner-state domain. 
The color code describes the minimum gap of the BdG spectra for all wave vectors.
\blue{The character of each phase can be identified by counting the boundary crossings 
along a vertical line, starting at zero magnetic field, i.e., topological or trivial 
for an odd or an even number of crossings, respectively.}
Along these boundaries the gap closes at $k=0$.  Starting from any point outside the
(A) All superconductor phases are equal to zero. 
(B) Phases are: 0 at one corner and $\pm \pi/6$ at the other corners, i.e. $\theta=\pi/6$ 
in Fig.\ \ref{Sample}(A).
(C) The same phase distribution, with $\theta=\pi/2$.
}
\label{Pdtri}
\end{figure}

For a symmetric triangle the corner states have equal probability
distribution at each corner \cite{Sitek15}, whereas in the presence of a
weak electric field $E$, here corresponding to 0.22 mV across the radius $R$,
they separate.  The wave functions still have some exponential tails along
the sides of the polygon, which are equivalent to the inter-chain hopping
introduced earlier.  The phase diagram shown in Fig.\ \ref{Pdtri}(A)
is obtained with a real valued superconductor gap $\Delta=0.5$ meV,
and can be compared with Fig.\ \ref{Fig1TT}(B) (where all $\phi_{\ell}=0$).
The fragmentation of the phase boundaries in three dark lines reflects
the presence of the three corners (edges) of the prismatic wire. The
boundaries approach each other when the aspect ratio of the triangle
($d/R$) decreases, which results in reduced overlap of the wave functions
of the corner states \cite{Manolescu17}.

The colors used indicate the minimum gap of the BdG
spectrum at any wave vector $k$, on a logarithmic scale, so the representation
is complementary to the two-color scheme of Fig.\ \ref{Fig1TT}(B) [or (A)].  Here
the topological phases can be identified by the number of crossings of the
dark lines. Along these lines the gap closes at $k=0$.  Starting from any point outside the
boundaries one enters into a topological Majorana phase after the first
intercept of a dark line, then into the trivial phase after the second
intercept, and again into the topological phase after the third intercept.

Next, in Fig.\ \ref{Pdtri}(B), we show the phase diagram obtained with a
complex valued superconductor gap, of constant modulus and variable phases,
which are zero at one corner and $\pm \pi/6$ at the other corners [i.e. $\theta=\pi/6$
in Fig.\ \ref{Sample}(A)].  We obtain a splitting (or anticrossing)
of the phase boundaries at the former crossing point, similar to that
shown in Fig.\ \ref{Fig5TT}(A), although now more pronounced than in 
the chain model.  

By further increasing the relative (angular) phase $\theta$ to $\pm \pi/2$ the 
boundaries of the quantum phase transitions become nearly parallel, Fig.\ \ref{Pdtri}(C).  
This result can be interpreted as an increased interaction between the 
corner states in the presence of the phase shift $\theta$ of the
superconductors.  Another consequence of this phase shift is that 
the absolute gap of the BdG spectrum decreases
in some topological regions, as indicated by the diffuse reddish regions,
suggesting that some topological states may become gapless.  This tendency is 
consistent with the results of the multiple chain model, compare Fig.\ \ref{Fig3TT}(B)
with Fig.\ \ref{Fig5TT}(C).

\blue{As with the coupled-chains model, we also tested the effect of using two 
superconductors with different gaps, for example by reducing the gap parameter $\Delta$ of one or
two superconductors by one half, and using no relative phase, $\theta=0$.  The
resulting phase diagrams were qualitatively like those shown in Fig.\
\ref{Pdtri}(B-C), although with lower energy gaps in the topological
phases. This indicates no particular gain by creating an asymmetry in this way,
compared to using the superconductors with the large gap and creating the 
asymmetry via the relative phase $\theta$. }

\begin{figure} 
\begin{center}
\includegraphics[width=6.0cm]{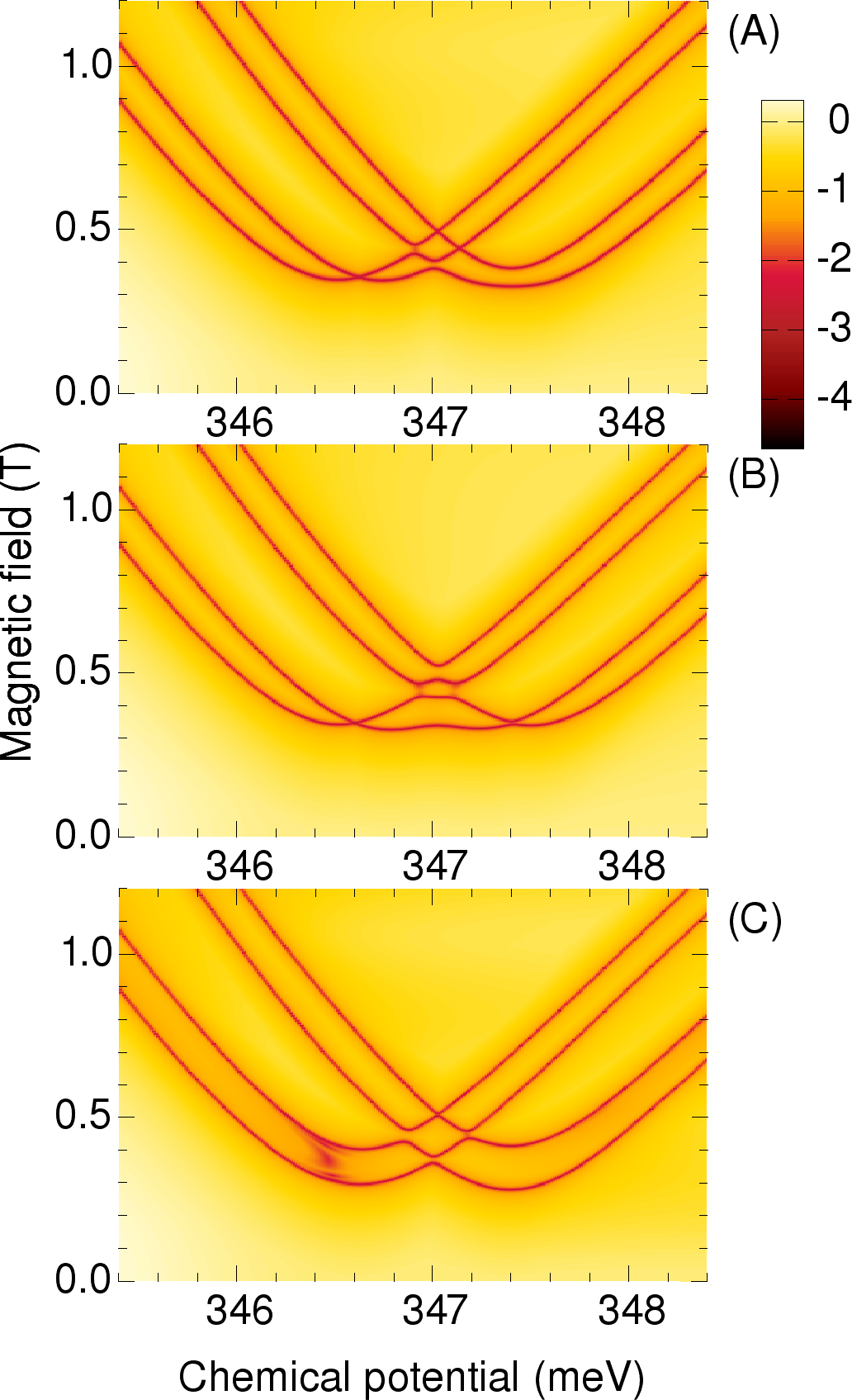}
\end{center}
\caption{Phase \blue{boundaries} for the square wire in the corner-state domain. 
The color code describes the minimum gap of the BdG spectra for all wave vectors.
\blue{The topological or trivial character of the phases can be identified by the 
number of boundary crossings, as described in the caption of Fig.\ \ref{Pdtri}.}
(A) The superconductor phases equal to zero. 
(B) The superconductor phases are zero and $\theta=\pi/2$, and the electric 
field perpendicular to the superconductors, see Fig.\ \ref{Sample}(B).
(C) Again $\theta=\pi/2$, but with the electric field parallel to the superconductors.
}
\label{Pdsqu}
\end{figure}

Finally, in Fig.\ \ref{Pdsqu} we show the phase diagrams obtained
with the geometric model for the square shell profile.  
\blue{Here, in the geometrical model, we use a particular setup for
the square geometry, with only two superconductors. Unlike in the coupled-chains 
model, in this case
the superconductors are also connected to the states localized on the
sides of the polygon, if those states would be populated, but this is not
the case for the chemical potentials used for Fig.\ \ref{Pdsqu}.}
First we note
that we obtain four phase boundaries, according to the presence of four
corner states.  As for the triangular geometry the trivial or topological
character of the phases is associated with odd or even number of boundary
crossings, respectively, when starting from the outer regions.  Therefore
the central zone of the phase diagrams is now topologically trivial.
In Fig.\ \ref{Pdsqu}(A) we show the results with $\theta=0$, i.e.
no phase shift between the superconductors [Fig.\ \ref{Sample}(B)]. The
electric field corresponds now to 60 mV per radius, and obviously 
the results do not depend on the two orientation considered here if $\theta=0$.

Remarkably, with a finite phase shift, here $\theta=\pi/2$, the phase diagrams
are different when the electric field is perpendicular, Fig.\ \ref{Pdsqu}(B), or parallel
to the superconductors, Fig.\ \ref{Pdsqu}(C), respectively. 
In the perpendicular case the phase frontiers are mostly changed in the central 
region, whereas in the parallel case they are more affected in the low field part.
In the first case the corner states with phase $\theta$ are separated energetically from 
those with zero phase, but they still interact when they are all grouped within or close to
the superconductor gap.  In the second case the states with the same superconductivity
phase are separated, and the frontiers tend to become parallel.


\section{Summary and conclusions}

In this work we have studied the phase diagram of core-shell nanowires
coupled with multiple parent superconductors using a simplified
tight-binding parallel-chain model. We found that applying a potential
that breaks the (intrinsic) rotation symmetry of the wire does not
modify the topology of the phase diagram, but  removes the gapless
superconducting phases that populate certain regions of the phase
diagram and partially stabilizes the topological superconducting
phase. Remarkably, finite phase differences between the parent
superconductors have dramatic effects. First, the  topology of the
phase diagram is modified. In particular the ``crossing points''
that characterize the phase diagram in the presence of a uniform
superconducting phase disappear and, upon increasing the Zeeman field,
we have an alternance of trivial and nontrivial phases for all values of
the chemical potential. More importantly, the low-field topological phase
becomes stable for a wide range of chemical potentials and the minimum
critical field $\Gamma_B^c$ can have arbitrarily low values.  We conclude
that by controlling the relative phases of the parent superconductors
coupled to the wire one can stabilize the topological superconducting
phase that hosts the zero-energy Majorana modes and  one  can obtain  a
powerful additional experimental knob for exploring a rich phase diagram
and  observing potentially interesting low-energy physics. Given the
potential experimental significance of these conclusions, we believe
that a more detailed and systematic investigation of these effects,
which is beyond the goal of the present work, would be warranted.

\blue{In particular, the effect of electrostatic interactions on the
properties of the normal electronic states in core-shell nanowires can
be important. The effect of interactions should be calculated using
a Schr\"odinger-Poisson scheme, e.g. like in Ref. \cite{Buscemi15},
to take into account both the interface potential between the core
and the shell, and the presence of the carrier density in the shell.
In addition, for Majorana devices, one should incorporate the effects due
to the presence of a parent superconductor, including the work function
difference between the superconductor and the semiconductor, as well as the
effects generated by gate-induced electric fields.  An efficient method
for implementing the Schd\"odinger-Poisson scheme in calculations using
realistic three-dimensional models of hybrid devices has
been recently proposed in \cite{Woods18}.  We emphasize that, due to the
corner and side localization, the electron-electron interactions have
nontrivial effects \cite{Sitek17}, which can modify the proximity-induced
superconductor gap and the phase diagram of the Majorana states
\cite{Gangadharaiah11,Sela11,Lutchyn11b,Stoudenmire11,Lobos12,Hassler12,Thomale13,Manolescu14}.
The calculation of the effective potential profile
is also essential for estimating the SOC in the nanowire.  Therefore,
accounting for the electrostatic effects represents a key step toward
a quantitative theory of Majorana physics in core-shell nanowires.}

\acknowledgments
This work was partially financed by the research funds of Reykjavik University and
by the Icelandic Research Fund.
TDS was supported in part by NSF DMR-1414683.


\bibliography{csmaj}

\end{document}